\documentclass[final,3p,times]{elsarticle}

\usepackage{amssymb}

\usepackage{graphicx,calc}
\usepackage{amsmath}
\usepackage{amsthm}
\usepackage{nccmath}
\usepackage{amssymb}
\usepackage{float}
\usepackage{caption}
\usepackage{subcaption}  
\usepackage[most]{tcolorbox}
\usepackage{bigints}
\usepackage[utf8]{inputenc}
\usepackage[T1]{fontenc}  
\usepackage{hyperref}
\usepackage{xcolor}
\usepackage{booktabs}
\usepackage{pgfplots}
\usepackage{float}
\usepackage{caption}
\usepackage{subcaption}  
\usepackage{amsmath}  
\usepackage{amsfonts}
\usepackage{amssymb}   
\usepackage{amsthm}
\usepackage{thmtools}
\usepackage{indentfirst} 
\usepackage{url}
\usepackage{hyperref}
\usepackage{wrapfig}
\usepackage{titlesec, color}
\usepackage{ifthen}
\usepackage{etoolbox}
\usepackage{fixmath}
\usepackage{eurosym}
\usepackage{bm}
\usepackage[ruled,vlined]{algorithm2e}
\usepackage{ mathrsfs }
\usepackage{multirow}
\usepackage{mathtools}

\theoremstyle{plain}
\theoremstyle{definition}

\usepackage{graphicx, color}
\graphicspath{{fig/}}

\usepackage{mathrsfs} 
\usepackage{lipsum}
\usepackage{algpseudocode}
\usepackage{xfrac}
\usepackage{nicefrac}
\DeclareUnicodeCharacter{2206}{\increment}

\newcommand{\Figure}{Figure}

\newcommand{\Tableau}{Table}

\newcommand{\Fig}[1]{\textbf{\Figure~\ref{#1}}}
\newcommand{\Tab}[1]{\textbf{\Tableau~\ref{#1}}}
\newcommand{\App}[1]{\textbf{~\ref{#1}}}

\newcommand{\Sec}[1]{\textbf{Section~\ref{#1}}}

\newcommand{\REV}[1]{{\textcolor{black}{#1}}}
\newcommand{\MOV}[1]{{\textcolor{black}{#1}}}

\journal{Journal}

\begin{document}

\begin{frontmatter}



\title{\REV{Normal contact of metainterfaces}: the roles of finite size and microcontact interactions
}


\author{Donald Zeka$^{a,b}$, Nawfal Blal$^b$, Fatima-Ezzahra Fekak$^{b,c}$, Arnaud Duval$^b$, Anthony Gravouil$^b$, Julien Scheibert$^d$}
 \affiliation{organization={Arts et Métiers Institute of Technology, CNRS, Bordeaux INP, Hesam Université, I2M, UMR 5295},
            city={Talence},
            postcode={F-33405},
            country={France}}

\affiliation{organization={INSA Lyon, CNRS, LaMCoS, UMR5259},
            city={Villeurbanne},
            postcode={69621}, 
            country={France}}
\affiliation{organization={Université Sidi Mohamed Ben Abdellah, ENSA, LISA},
            city={Fez},
            postcode={30000}, 
            country={Morocco}}
  \affiliation{organization={CNRS, Ecole Centrale de Lyon, ENTPE, LTDS, UMR5513},
            city={Ecully},
            postcode={69134},
            country={France}}

\begin{abstract}
The design of contact interfaces that meet quantitatively a specified friction law (friction force vs normal force) is challenging due to the multi-scale and multi-physics nature of contact interactions. Recently, a concept was proposed to address this question in the case of dry elastic microarchitected contact interfaces, so-called metainterfaces. These take their macroscopic friction properties from an array of discrete asperities whose geometrical descriptors are optimized through an inverse design phase. \REV{Such design is based on the experimentally-observed proportionality between friction force and real contact area under pure compression, reducing the friction problem to a simpler contact mechanics problem of designing the contact area. In this context, the design strategy assumes} that asperities are placed on a linear elastic half-space and behave independently from each other. Both assumptions are likely to fail in experimental realizations of metainterfaces, potentially inducing discrepancies between the actual and target behaviours. Here, we use full 3D finite element modelling to critically assess the validity of those \REV{two} assumptions in existing experimental metainterfaces, and their potential impact on the design quality. The results first confirm the validity of the strategy, in the conditions in which it was used in the literature. Then, by systematically varying the spatial arrangement of asperities, their interdistance and the size of their elastic base, we identify conditions under which the literature assumptions fail. Our findings provide critical
insights into the robustness and practical limitations of the metainterface design strategy and guidelines for its future improvements.
\end{abstract}



\end{frontmatter}


\section{Introduction}
\label{sec:intro}

Friction, the force that resists relative motion between contacting surfaces, is an everyday physical phenomenon with significant implications across engineering, manufacturing, and daily life. Friction control is crucial in advanced technologies such as precision robotics, where inadequate frictional regulation can lead to positioning inaccuracies, stick-slip motion, and overall performance degradation \cite{bona2005friction}. Despite its fundamental nature, the reliable prediction and precise control of dry friction (i.e., in the absence of lubricants), remains an open challenge. Dry friction emerges from a complex interplay of mechanisms that are highly sensitive to material properties, loading conditions and the microscopic topography of the contacting surfaces \cite{vakis2018modeling, sahli2019shear}. 

The multiscale, random nature of the roughness of most real surfaces, coupled with the potential impacts of material deformation, surface energy and their heterogeneities, makes the predictive modeling of friction from first principles challenging (see, e.g., \cite{xu2022asperity}). Consequently, the design of systems where dry friction is a dominant factor, such as robotic grippers \cite{liu2023beyond}, haptic interfaces \cite{basdogan2020review} and sports equipment \cite{worobets2015influence}, often relies on iterative, time-consuming, and costly empirical testing. This dependence on trial-and-error restricts the ability to rapidly design and optimize such technologies.

Traditional surface engineering approaches to control friction typically involve empirical techniques like the application of specialized coatings or texturing  at different scales \cite{costa2022tailored, wang2023performance}. While these methods have yielded success in specific applications (see, e.g.,~\cite{scheibert2009role, lamping2022frictional} for bioinspired textures), they lack a comprehensive theoretical framework. We still miss a systematic, generalizable methodology to translate any desired, macroscopic frictional response into a deterministic design for surface topography and/or surface coating. Bridging this gap between a functional specification and a manufacturable surface design would represent a significant breakthrough in tribology and contact mechanics, and in all application fields where functional interfaces are central.
In a recent first step in this direction, a rational strategy for designing elastic multicontact interfaces with tailored dry frictional properties was proposed in \cite{aymard2024designing}. This approach circumvents the intricacies of modeling natural rough surfaces by instead engineering simplified microarchitected interfaces, so called frictional metainterfaces in analogy with the field of metamaterials \cite{craster2023mechanical, mariano2022homogenization}. Metainterfaces arise from the contact between a smooth, rigid surface and an elastic surface decorated by a carefully engineered population of discrete microasperities. By fixing the material pair, the design problem reduces to prescribing the topography of the elastic surface, i.e., the geometrical properties of each individual asperity, such as its shape, height with respect to the mean plane, and in-plane location. The vast combinatorial space of possible multi-asperity topographies gives rise to a rich spectrum of potential macroscopic frictional behaviors (including ones not found in nature, see~\cite{aymard2024designing}), opening a path to systematically "program" friction by engineering the contact interface.

In this design strategy, a target friction law, i.e., the relationship between the friction force, $F$, and the normal load, $P$, is used as input. Suitable topographies are then identified through the inversion of a multicontact friction model, fed by the experimental calibration of the compression and shear behaviour of the individual asperities. In the examples of application of this strategy that were presented in~\cite{aymard2024designing}, the authors used 64 spherical asperities, all with the same material (polydimethylsiloxane, PDMS) and the same radius of curvature, $R$, organized along a $8\times8$ square lattice of pitch $d$. The friction force $F$  was found to be directly proportional to the real contact area under pure compression, $A_0$, so that a specification on $F(P)$ is equivalent to a specification on $A_0(P)$, reducing the friction problem to a simpler contact mechanics problem. The contact model assumed Hertzian behaviour \cite{barber2018contact} for the compression of each microcontact and elastic independence of the asperities. This latter assumption implies that the exact in-plane locations of the asperities are irrelevant. $A_0$ and $P$ were thus simple sums of 64 Hertz contacts:
\begin{align}
A_0(\REV{Z})&=\sum_{i=1}^{64} a_{0,i}(\REV{Z}) = \pi R \sum_{i=1}^{64} (h_i - \REV{Z}) \REV{\Theta}(h_i - \REV{Z}),\label{Eq:A}\\
P(\REV{Z})&=\sum_{i=1}^{64} p_i(\REV{Z}) = \frac{4}{3}E^*\sqrt{R} \sum_{i=1}^{64} (h_i - \REV{Z})^{3/2} \REV{\Theta}(h_i - \REV{Z}),\label{Eq:P}
\end{align}\label{Eq:GW}
\noindent where $a_{0,i}$ and $p_i$ are the contact area and normal force of asperity $i$, \REV{$Z$} is the altitude of the rigid indenting plane (the \REV{arbitrary} reference is the same as for the asperity heights $h_i$), $E^*=\frac{E}{1-\nu^2}$ with $E$ and $\nu$ the Young modulus and Poisson ratio of the elastic material), and the Heaviside function $\Theta$ ensures that asperities such that $h_i<\delta$ (out-of-contact asperities) have vanishing contributions to $A_0$ and $P$.  Note that the model of Eqs.~\ref{Eq:A} and~\ref{Eq:P} can be seen as a discrete version of the Greenwood-Williamson model~\cite{greenwood1966contact}. In these conditions, the output of the inverse design is a list of asperity heights, $h_{i\in[1:64]}$, that satisfies a given set of specifications on the compression law, $A_0(P)$.

While the model of Eqs.~\ref{Eq:A} and~\ref{Eq:P} provides a simplified and effective foundation for the design process, it is predicated on several assumptions that may limit its practical robustness. 
\REV{One assumption is the already mentioned proportionality between the friction force and contact area. Although such a relationship is not universal (counter-examples relate, e.g., to adhesion~\cite{berman1998amontons}, capillarity~\cite{hsia2021rougher} or plastic deformation~\cite{liang2021experimental}), it was used in~\cite{aymard2024designing} to capture the calibrated individual asperity behaviour, consistent with repeated literature observations in diverse contact conditions (see e.g.~\cite{berman1998amontons,enachescu1999observation,degrandi2012sliding,sahli2018evolution, mergel2019continuum,liang2021experimental}). In the present modelling work, we accept this experimentally-based proportionality between $F$ and $A_0$ and focus on how $A_0$ is controlled by the list of asperity heights. We will thus only consider the compression law $A_0(P)$ of metainterfaces, the friction law $F(P)$ being assumed to be directly proportional to it.}

In this article, we critically assess two \REV{additional} assumptions, which are potentially invalid in experimental realizations of metainterfaces. First, using Hertz's model implicitly assumes linear elastic half-spaces and small indentations compared to the radius of curvature $R$. In contrast, the metainterfaces of~\cite{aymard2024designing} involved a finite sized sample (finite depth and lateral dimensions) and indentations as large as 0.4$R$. Second, the asperities are assumed to be independent, but elastic interactions are fundamentally long-ranged \cite{ciavarella2006re, afferrante2012interacting}, and thus exist whatever the interasperity distance $d$.

Here, our strategy to examine the validity of those assumptions consists in building a finite elements model of the full geometry of the experimental metainterfaces used in~\cite{aymard2024designing}. By taking explicitly into account the finite size and asperity-based topography of the textured samples, we can compare quantitatively the simulated and experimentally measured compression laws, $A_0(P)$, and identify the origin of potential discrepancies. The model is then further used to vary systematically various system parameters that are not explicitly design parameters, but are expected to affect the metainterface response: the interasperity distance $d$, the relative locations of the asperities along the plane and the dimensions of the asperity-bearing elastic block.

The article is organised as follows. The proposed finite element model is described in \Sec{sec:methods}. \Sec{sec:single} focuses on the non-linear behaviour of individual asperities. \Sec{sec:directsimu} presents the direct comparison between the simulated and experimental responses of all three types of metainterfaces shown in~\cite{aymard2024designing}. \Sec{sec:interactions} explores the effects of elastic interactions between microcontacts by shuffling the locations of asperities on the plane and changing the inter-asperity distance. In \Sec{sec:FSE} are investigated finite size effects of the textured sample, both its lateral size and thickness. \REV{A discussion and a conclusion are given in \Sec{sec:discussion} and \Sec{sec:conclusions}, respectively.}

\section{Finite element model}
\label{sec:methods}

\begin{figure}[ht!]
\centering
\begin{subfigure}[b]{0.4\linewidth}
    \includegraphics[width=\linewidth]{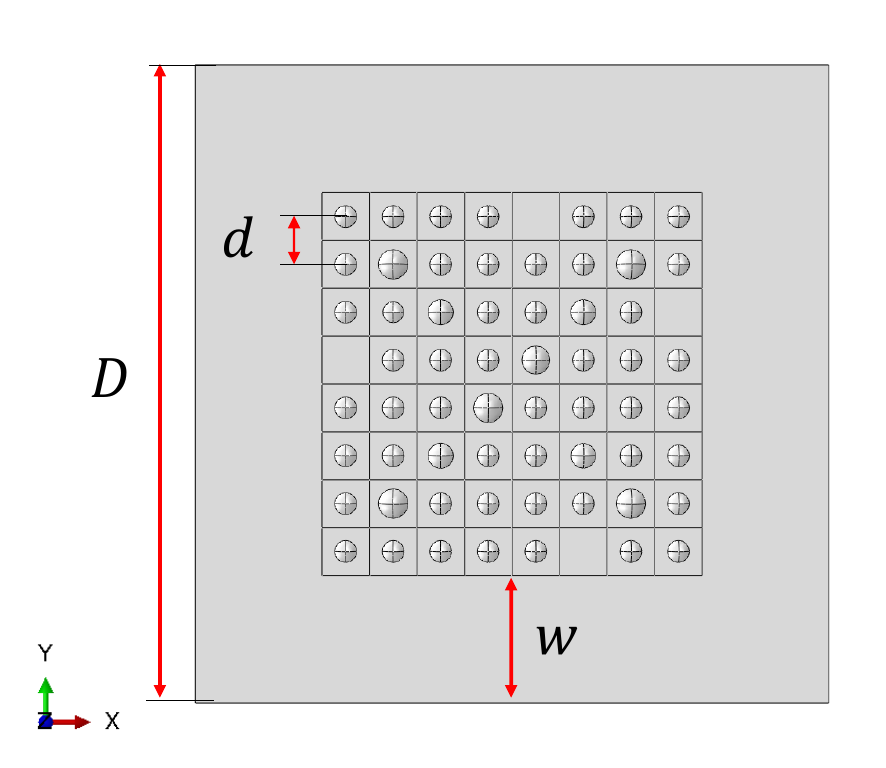}
    \caption{Geometric model, top view.}
   \label{}
\end{subfigure}
\hspace{1.5cm}
\begin{subfigure}[b]{0.4\linewidth}
    \includegraphics[width=\linewidth]{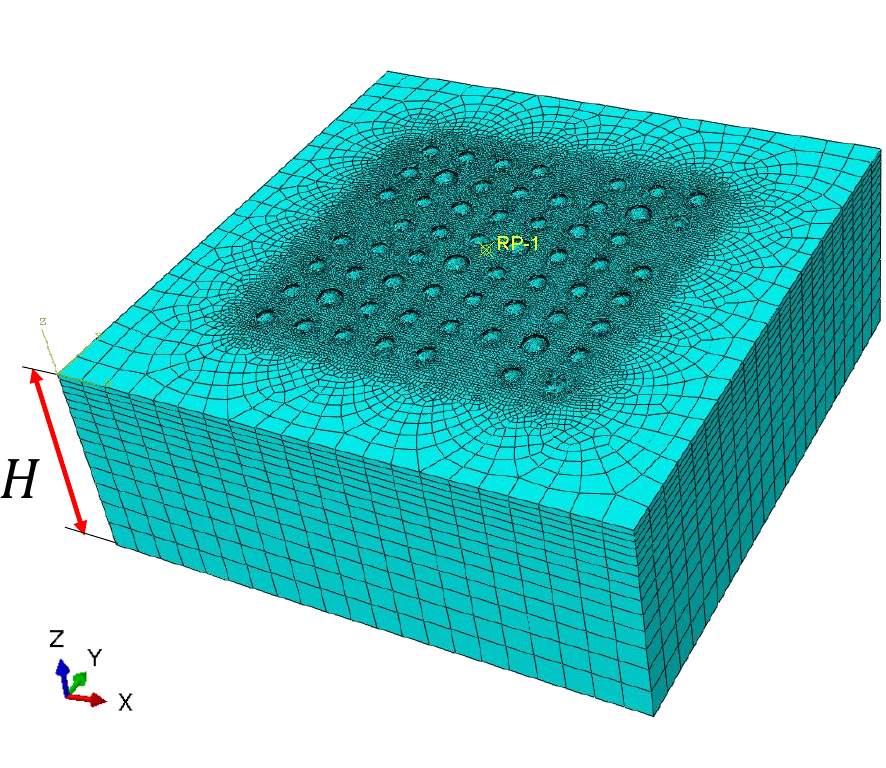}
    \caption{Finite element model, 3D view.}
    \label{}
\end{subfigure}
\begin{subfigure}[b]{0.25\linewidth}
    \includegraphics[width=\linewidth]{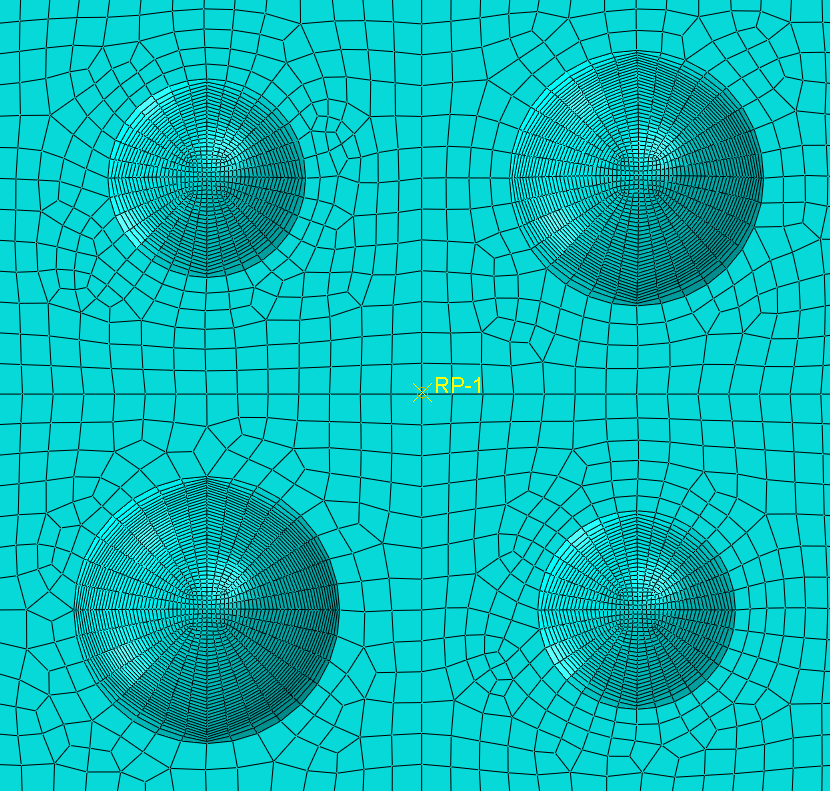}
    \caption{Detail on four central asperities.}
   \label{}
\end{subfigure}
\caption{Geometry and finite element model of a typical metainterface (the one with asperity heights given in \Tab{Tab:HOP}). (a): Geometric model, top view. (b): Finite element model, 3D view. (c): Detail on the four central asperities.}
\label{fig:model}
\end{figure}

The finite element (FE) model is constructed in ABAQUS/Standard~\cite{abaqus2011abaqus} with the exact geometry of the experimental, silicone elastomer metainterfaces used in~\cite{aymard2024designing}. As shown in \Fig{fig:model}, the square top surface ($D \times D$, with $D$=20\,mm) of a parallelepipedic elastic base of thickness $H$=7.2\,mm is decorated by an 8 $\times$ 8 square lattice of asperities, with a pitch of $d$=1.5\,mm in both the $x$ and $y$ directions. Each asperity is a spherical cap with a radius of curvature $R$=526\,$\mu$m, and with a summit that emerges by a height $h_i$ from the otherwise flat and smooth surface. The {12\,mm\,$\times$\,12\,mm textured square ($8\times1.5$\,mm=12\,mm) is located at the center of the top surface. For the textures, we consider five reference lists of asperity heights, corresponding to the five metainterfaces used in~\cite{aymard2024designing} as illustrations of their design strategy. Those lists are given in \App{sec:appendixA}. \Fig{fig:model} shows a typical finite element model of a metainterface, the one corresponding to the aperity heights of \Tab{Tab:HOP} (metainterface with a friction law passing through three predefined operating points).

The material model is assumed to be Neo-Hookean, a behaviour previously shown to capture non-trivial contact mechanics behaviours of PDMS (see for instance~\cite{lengiewicz2020finite, mergel2019continuum, mergel2021contact, nguyen2011surface}). The elastic parameters are taken from~\cite{aymard2024designing}: the PDMS is classically assumed incompressible ($\nu=0.5$), and its Young modulus is either $E$=1.02\,MPa or 1.14\,MPa, depending on the sample (due to two different batches of PDMS used in~\cite{aymard2024designing}), leading to reduced moduli $E^*$=$\frac{E}{1-\nu^2}$=1.36\,MPa or 1.52\,MPa, respectively.

The boundary conditions and the loading sequence also correspond to the experimental ones. The displacements are fully constrained at the bottom surface of the elastic base. The asperity-bearing surface is indented by a rigid horizontal plane (parallel to the $x-y$ plane), kinematically constrained in all directions except for the vertical displacement ($z$-direction), along which a normal load $P$ is applied to the plane. 

Eight-node hexaedral C3D8H elements are used to discretize the model, with the hybrid formulation to take into account the incompressibility constraint. The augmented Lagrangian method \cite{wriggers2006computational, alart1991mixed, pietrzak1999large} is used to enforce the normal contact conditions with a surface-to-surface approach. A frictionless behavior is adopted for the tangential contact, \REV{an assumption with negligible impact on the compression behavior, as shown in~\Sec{sec:single}}. A reference element size of $0.015$\,mm is used on the spherical caps and $0.1$\,mm for the textured square. This discretization has been chosen to keep tractable computational cost and memory usage, while capturing well the main physical phenomena occurring at metainterfaces. \REV{Indeed, when decreasing the reference element size of the spheres to 0.01\,mm, the relative root mean square difference in total contact area (over the 64 microcontacts) is only about 1.2\%. \Sec{sec:single} contains more information about mesh sensitivity at the single asperity level}. Note that those simulations are actually challenging, because of the large number of degrees of freedom (between three and four millions due to 64 finely meshed asperities) and the multiple sources of nonlinearities (including a large number of localized contacts, incompressibility, and large strains).

\section{Single microcontact behaviour }
\label{sec:single}

We first examine the compression behaviour of a single microcontact between a surface asperity and the rigid plane. In the design strategy of metainterfaces, this behaviour represents the building block from which the desired macroscale behaviour can be constructed. In~\cite{aymard2024designing}, the single asperity behaviour was observed to be well-captured experimentally by Hertz's model. This is a priori unexpected because most assumptions of Hertz are violated: the asperities are not parabolic, they are not perfectly smooth, they are not frictionless, their material behaviour is not linear elastic \cite{nguyen2011surface} and they are submitted to indentations $\delta$ \REV{(the vertical displacement of the rigid plane relative to the summit of the highest asperity in its underformed state)} that may reach 40\% of the asperity's curvature radius, $R$.

\Fig{fig:onesphere} shows the evolutions of the contact area, $a_0$ (extracted using the CAREA output of Abaqus), and the indentation \REV{ratio, $\delta/R$}, as functions of the imposed normal force, $p$, for a single microcontact. The explored range of indentation goes up to almost $\delta/R=0.4$, a value reached experimentally on some metainterfaces. To test the suitability of the reference element size of 0.015\,mm used for full metainterface calculations, we \REV{performed a sensitivity analysis on a single microcontact. We considered a single spherical cap mounted at the center of the same elastic base of size 20\,mm\,$\times$\,20\,mm\,$\times$\,7.2\,mm, and three different discretizations, with an element size of either 0.015\,mm, 0.01\,mm or 0.005\,mm. The resulting curves for the contact area, $a_0$, are overplotted on \Fig{fig:onesphere_mesh}. The relative root mean square variation in contact area is 1.4\% when reducing the mesh size from 0.015\,mm to 0.01\,mm, and 1.2\% more when further reducing to 0.005\,mm.}
\REV{These variations are sufficiently small} to consider that contact calculations are converged when using the reference element size of 0.015\,mm (\REV{the size that is} used in all the rest of the article \REV{for full metainterface computations}).

\begin{figure}[ht!]
    \centering
    \includegraphics[width=0.45\linewidth]{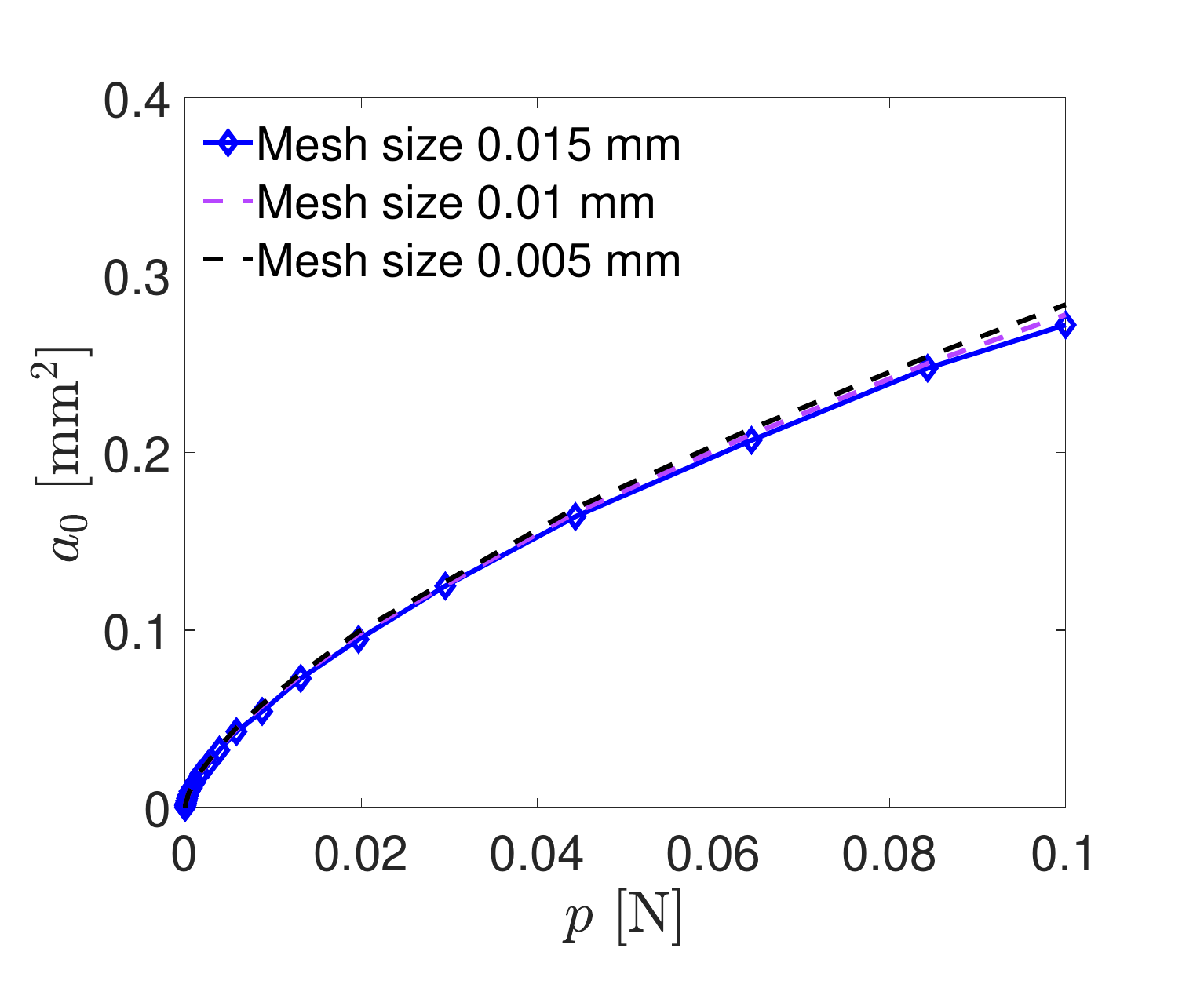}
    \caption{\REV{Sensitivity to the mesh size of a single microcontact. Evolution of the contact area, $a_0$, as a function of the normal force, $p$, for three values of the reference mesh size (see legend).}}
    \label{fig:onesphere_mesh}
\end{figure}

\begin{figure}[ht!]
\centering
\begin{subfigure}[b]{0.45\linewidth}
\includegraphics[width=\linewidth]{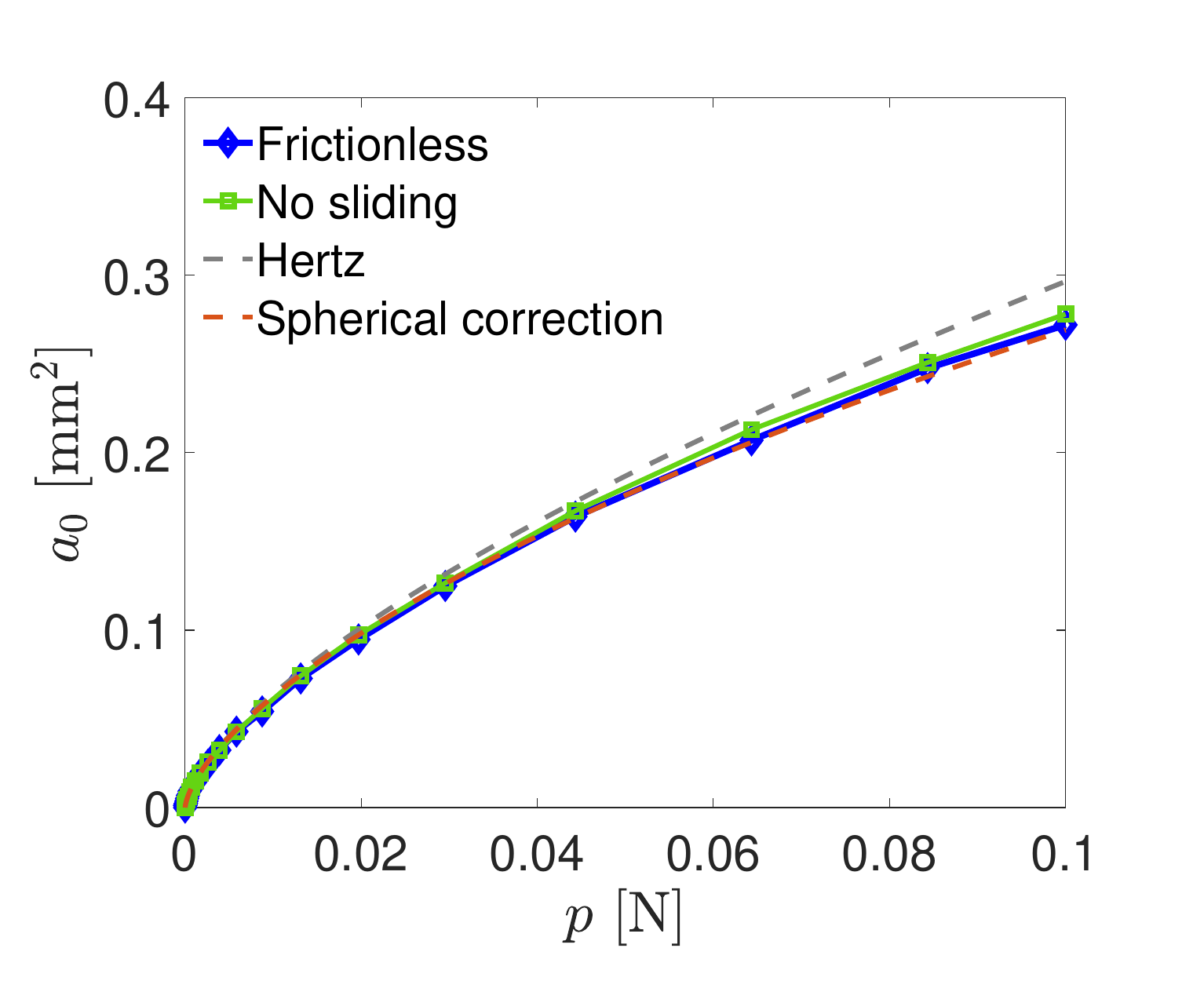}
\caption{$a_0$ as a function of $p$.}
\end{subfigure}
\begin{subfigure}[b]{0.45\linewidth}
\includegraphics[width=\linewidth]{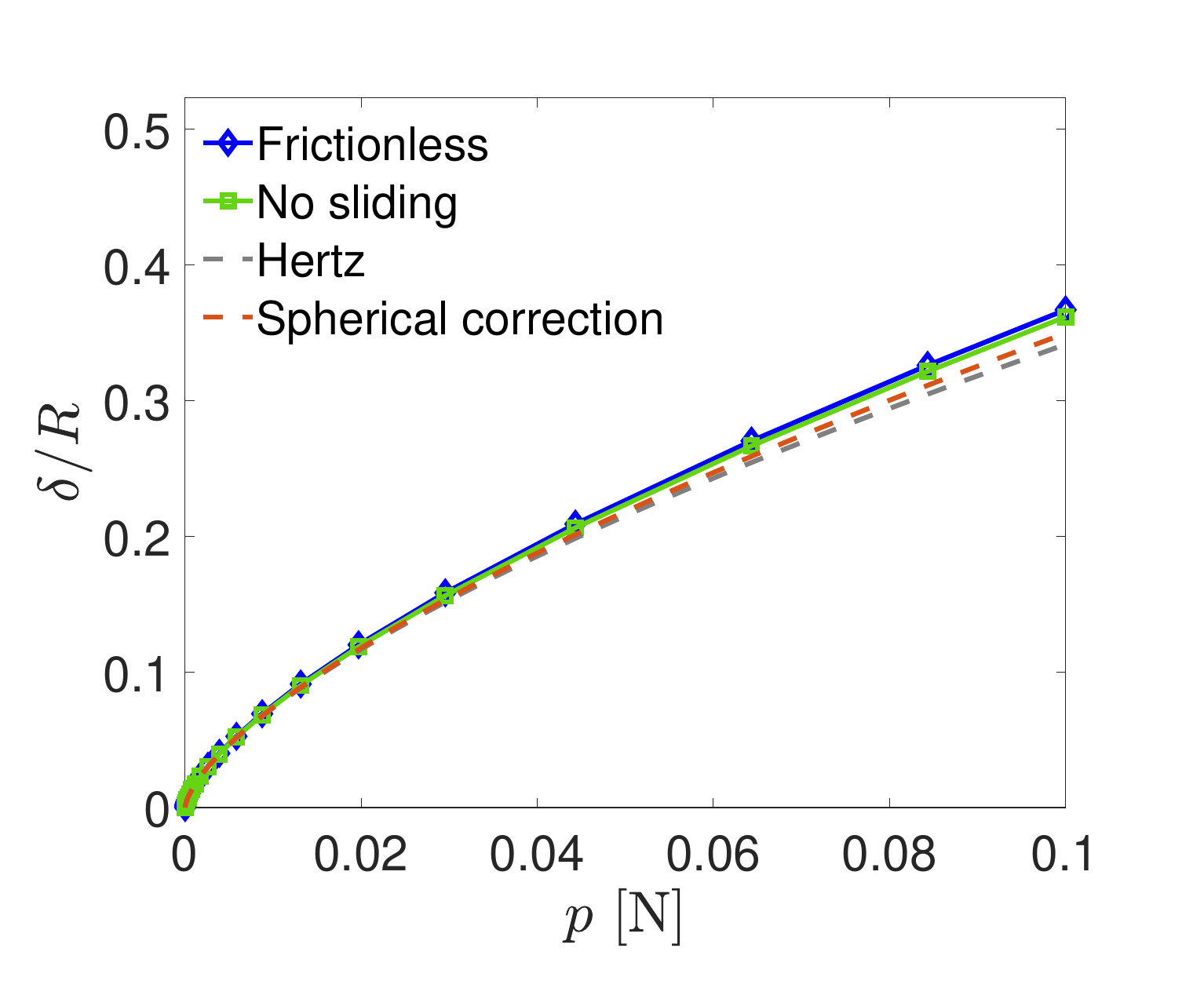}
\caption{$\delta\REV{/R}$ as a function of $p$.}
\end{subfigure}
\caption{\REV{Comparison between the proposed FE numerical simulations (frictionless conditions, solid blue lines and diamonds) for a single microcontact, and three other models: the same FE calculations but with no sliding interfacial conditions (solid green line and squares),  Hertz (dashed grey lines) and the spherical correction of Hertz by Segedin~\citep{segedin1957relation} (dashed red lines). (a) $a_0$ and (b) $\delta/R$ as functions of $p$.}}
\label{fig:onesphere}
\end{figure}
\REV{To test the impact of the frictionless assumption at the interface, we performed a computation in the other limit of no sliding for the tangential behavior. The resulting behavior, plotted in \Fig{fig:onesphere}, overall differs from the frictionless case by a relative root mean square variation of 1.4\%. Given that the real tangential contact conditions lie between these two extreme cases, it indicates that those contact conditions have negligible influence on the compression behavior, as also shown in \cite{mu2025nonlinear}}.

We now focus on the compression behaviour \REV{shown in \Fig{fig:onesphere}.} As expected, at small indentations (up to $\delta/R$$\sim$0.1, corresponding to $p$$\sim$0.015\,N), the observed behaviour is Hertzian. Beyond, the behaviour deviates from Hertz, with the contact area $a_0$ being smaller and the indentation $\delta$ being larger than those predicted by Hertz, for the same imposed normal force, $p$. In \Fig{fig:onesphere}, we compare our FE calculation (solid blue line) not only to Hertz (dashed grey line) but also to Segedin's model~\citep{segedin1957relation} (dashed red line). This latter model adopts the same assumptions as Hertz, except that the asperity's shape is truly spherical, and not parabolic as in Hertz. Remarkably, for the contact area, which is the quantity that ultimately controls friction, our calculations closely follow Segedin's prediction, suggesting that the observed discrepancy with Hertz is mainly due to the spherical shape of the asperity. The residual discrepancies between FE and Segedin seen when plotting $\delta$ vs $p$ are presumably due to large strain effects such as those discussed in~\cite{mu2025nonlinear}, and not taken into account in Segedin's model.

\section{Simulation of literature metainterfaces}\label{sec:directsimu}

In this section, we consider five reference simulations corresponding to the five metainterfaces tested experimentally in~\cite{aymard2024designing}. Two of them correspond to quasilinear laws (denoted as QL1 and QL2 below), two to bilinear laws (BL1 and BL2) and one to a law passing through three operating points (OP). The location and heights of the asperities follow that given in \App{sec:appendixA} (the asperity index denotes positions within the $8\times8$ square lattice ordered from top left to bottom right). The dimensions and material properties of the elastic base are the ones provided in \Sec{sec:methods}.

For each case, we compute the evolution of the total contact area, $A_0$, as a function of the imposed normal force, $P$, within the same range as that explored experimentally. We then compare this curve with both the as-designed curve based on the assumption of independent Hertz contacts (Eqs.~\ref{Eq:A} and~\ref{Eq:P}) and the experimental curve reported in~\cite{aymard2024designing}. To perform such comparison, we adopt the rescaling used in the figures of~\cite{aymard2024designing} and plot the contact area, $A_0$ (which is directly equal to their ratio of the friction force over an experimentally calibrated proportionality factor), versus the rescaled normal force, $P/E^*$. 

\begin{figure}[ht!]
\centering
\begin{subfigure}[b]{0.425\linewidth}
    \includegraphics[width=\linewidth]{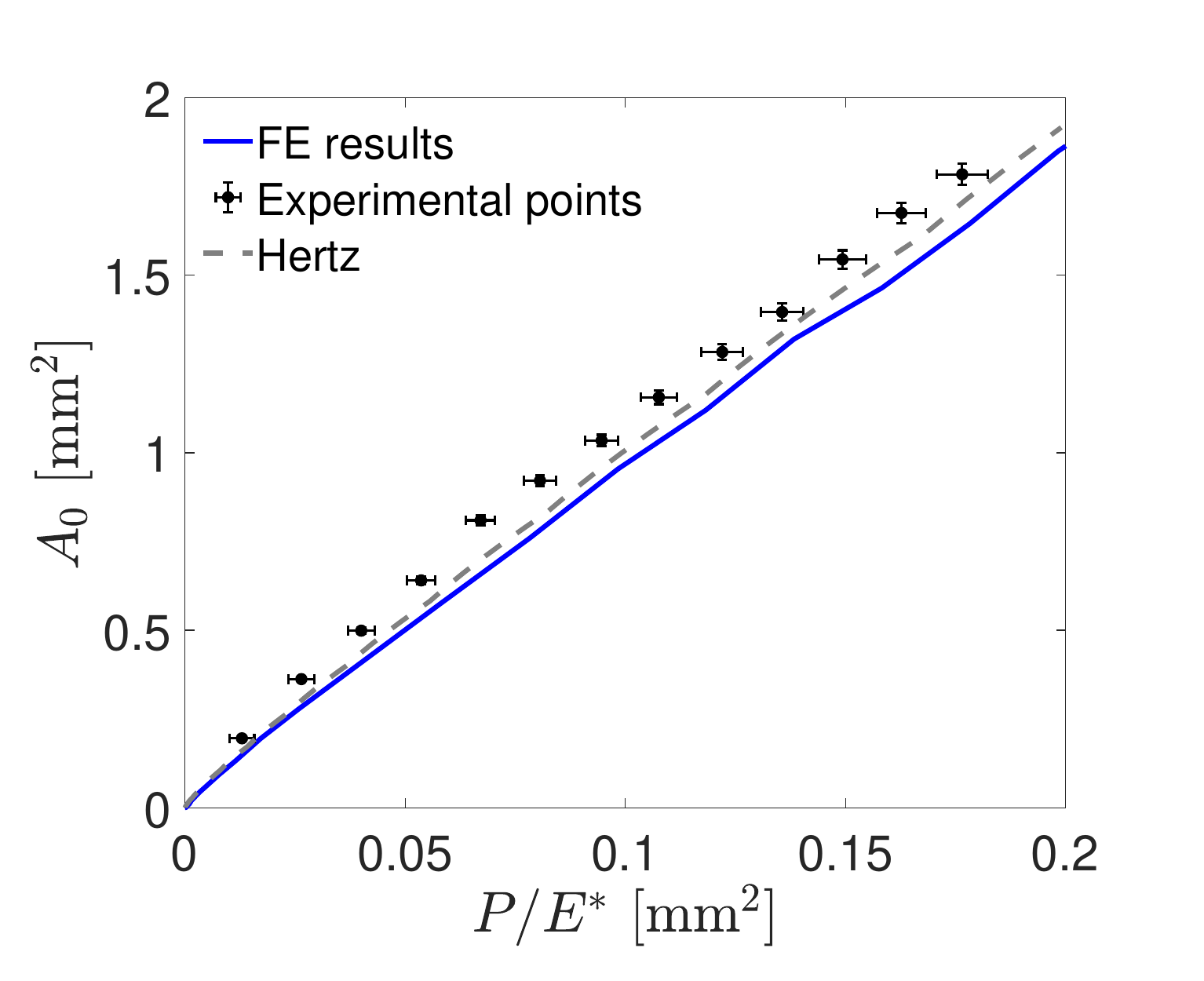}
    \caption{Quasilinear law 1 (QL1)}
    \label{}
\end{subfigure}
\begin{subfigure}[b]{0.425\linewidth}
    \includegraphics[width=\linewidth]{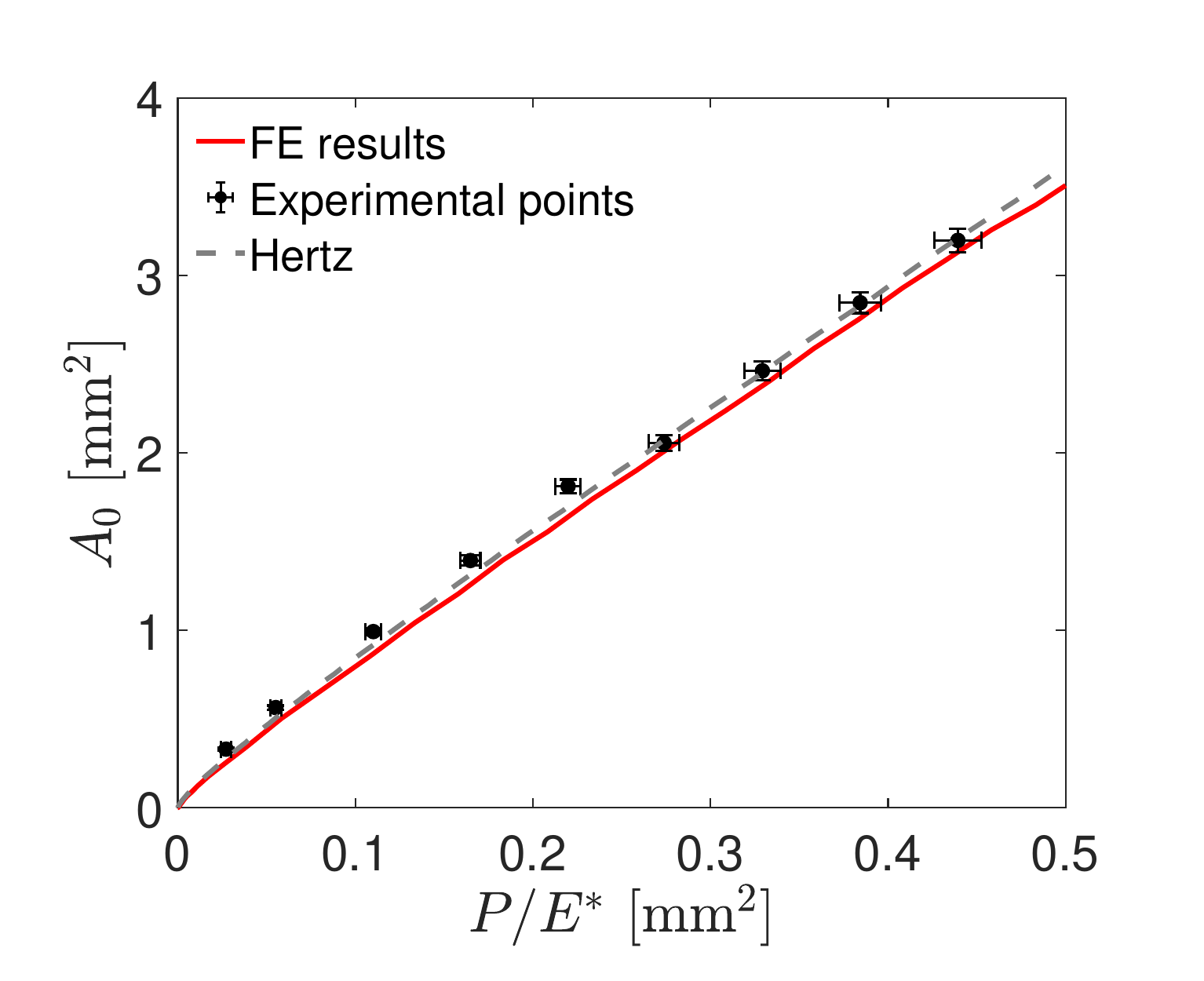}
    \caption{Quasilinear law 2 (QL2)}
   \label{}
\end{subfigure}
\begin{subfigure}[b]{0.425\linewidth}
    \includegraphics[width=\linewidth]{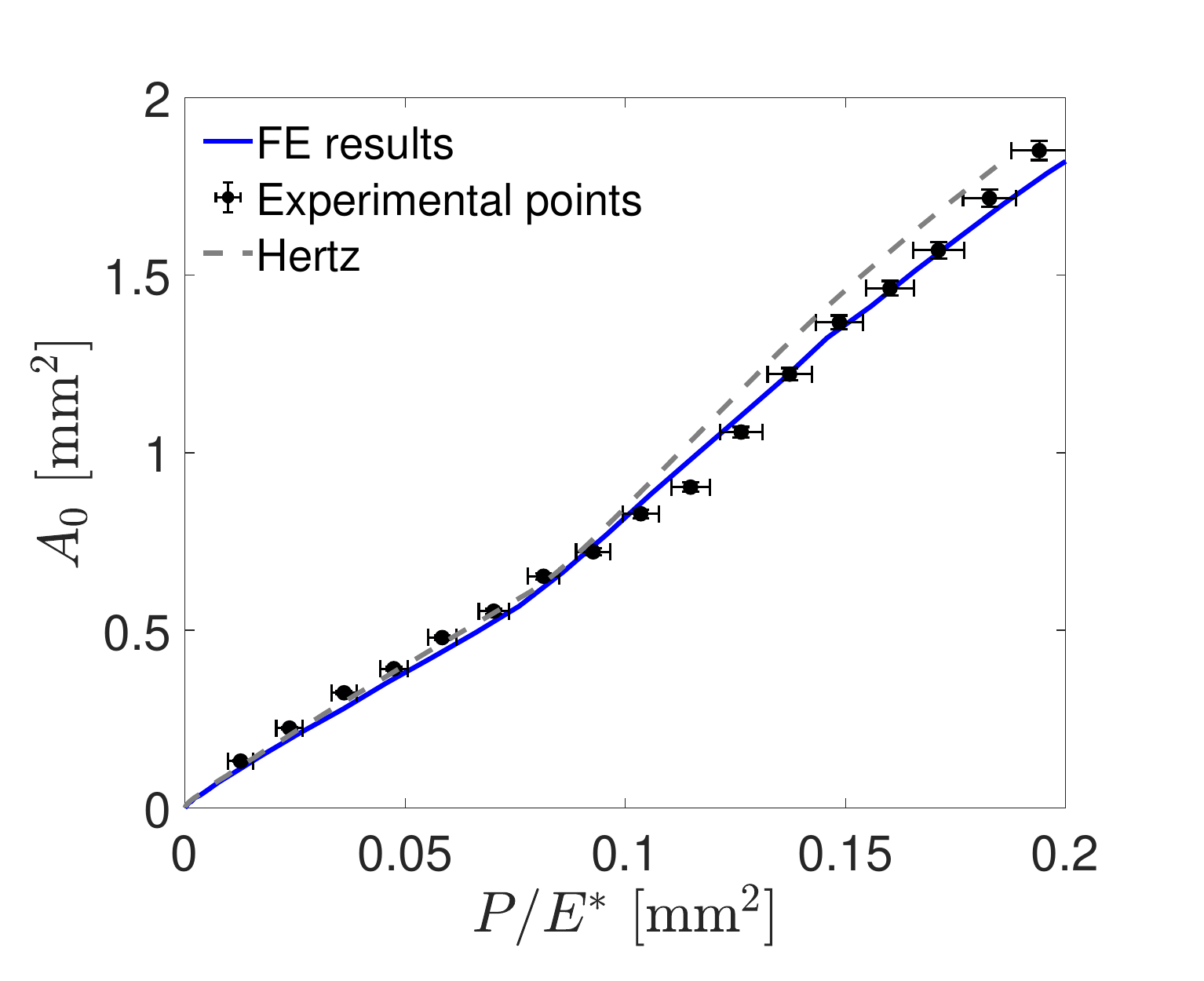}
    \caption{Bilinear law 1 (BL1)}
   \label{}
\end{subfigure}
\begin{subfigure}[b]{0.425\linewidth}
    \includegraphics[width=\linewidth]{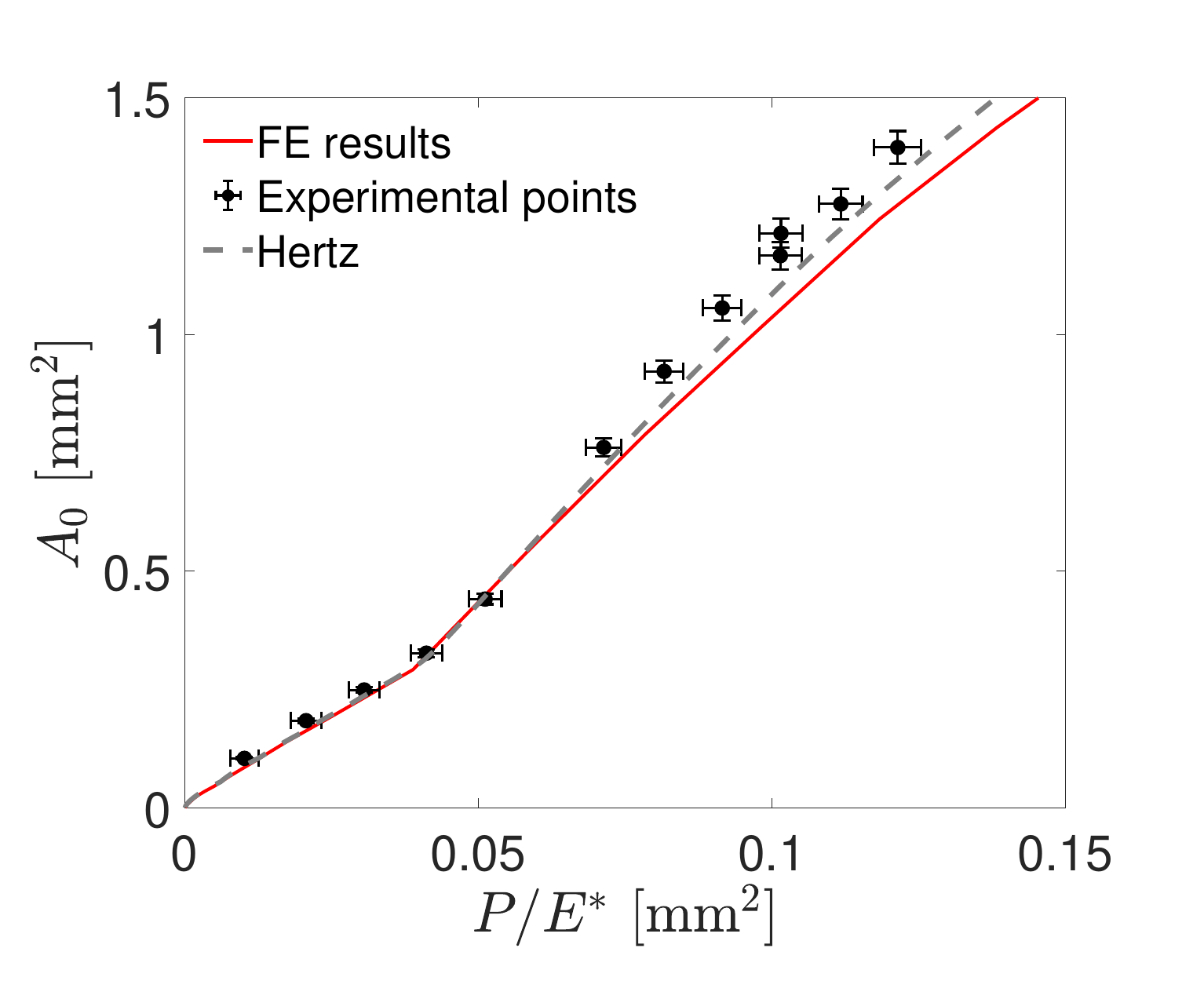}
    \caption{Bilinear law 2 (BL2)}
   \label{}
\end{subfigure}
\begin{subfigure}[b]{0.425\linewidth}
    \includegraphics[width=\linewidth]{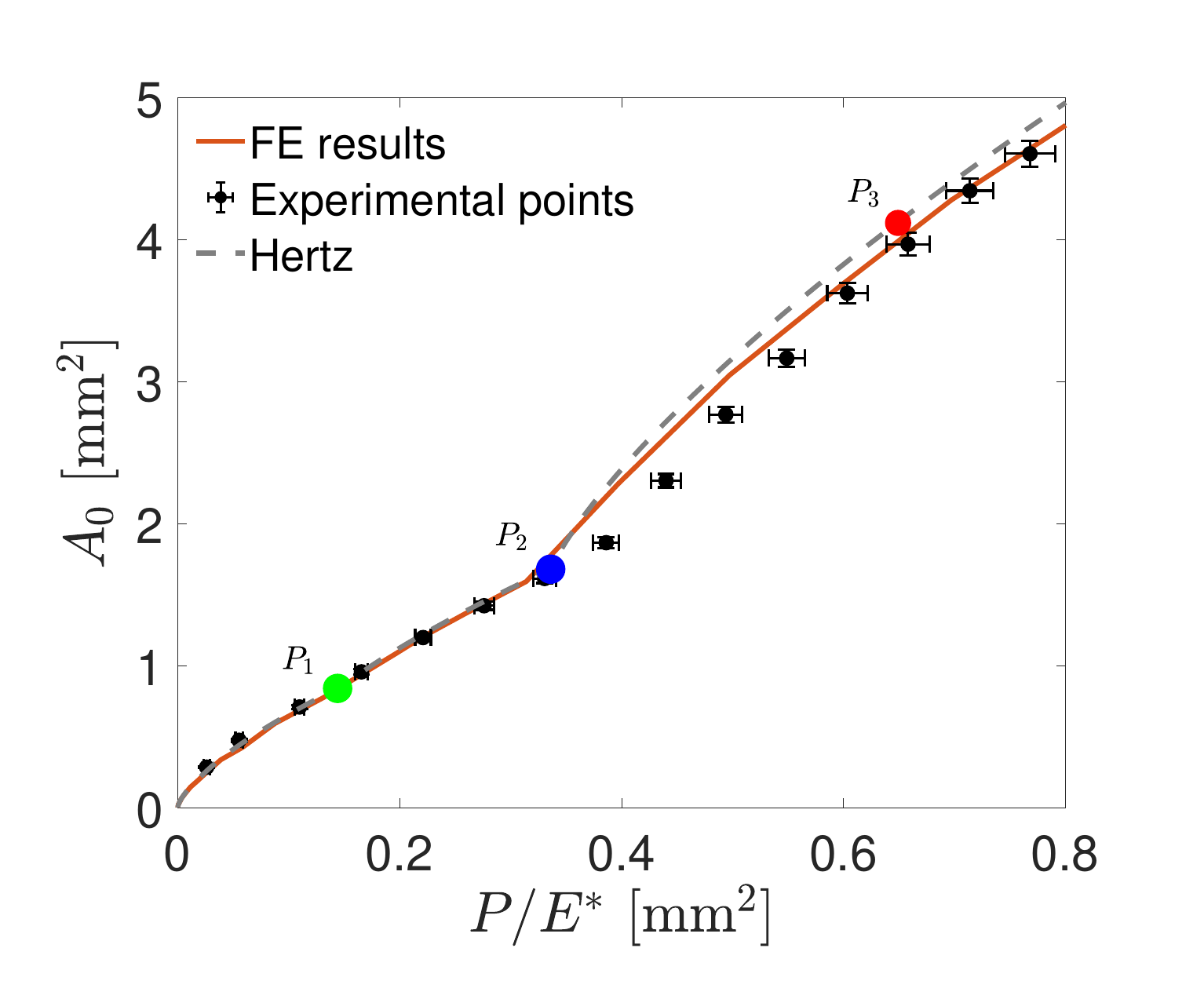}
    \caption{Operating points law (OP)}
   \label{}
\end{subfigure}
\caption{Direct simulations of the evolution of the contact area, $A_0$, vs the rescaled normal force, $P/E^*$, for the five metainterfaces tested experimentally in~\cite{aymard2024designing}. Black symbols are the experimental points. Dashed lines are the design predictions based on independent Hertz contacts. Solid colored lines are the FE simulations. (a) and (b): quasi-linear (QL) laws whose slope relates to the metainterface's friction coefficient. (c) and (d): bilinear (BL) laws with similar first and second slopes, but different crossover between the slopes. (e): law passing through three predefined operating points (OP). The green, blue, and red discs are the target operating points.} 
\label{fig:DirectSimu}
\end{figure}

\Fig{fig:DirectSimu} shows such comparisons for all five metainterfaces. First, the simulations do capture very well both the shapes and amplitudes of all experimental curves, without any adjustable parameter. This global quantitative agreement indicates that our FE calculations include all the dominating physical ingredients that dictate the compression behaviour of the metainterfaces reported in the literature.

Second, the FE results are, globally, also close to the prediction based on Eqs.~\ref{Eq:A} and~\ref{Eq:P}, showing that the underlying assumptions are essentially valid in the reference conditions tested here. This good agreement actually validates the overall design strategy developed in~\cite{aymard2024designing}.

\REV{In this context, accurate designs can be reached without resorting to our FE model. However, it is a unique tool to explore mechanical phenomena that may not have expressed themselves significantly in literature metainterfaces, but may do so when changing system parameters. This is why,} in the rest of the article, we will assess the role, on the law $A_0(P)$ (denoted below as compression law), of system parameters that do not appear in  Eqs.~\ref{Eq:A} and~\ref{Eq:P}, and are thus not accounted for in the design of the five metainterfaces of \Fig{fig:DirectSimu}. In the experiments, those parameters were chosen based on intuition and/or experimental constraints. They include the particular locations of the different asperities on the 64 nodes of the square lattice, the interdistance between asperities, and the sizes of the elastic base.

\section{Effect of elastic interactions between microcontacts}
\label{sec:interactions}

An important physical assumption underlying the multicontact model of~\cite{aymard2024designing} (Eqs.~\ref{Eq:A}--\ref{Eq:P}) is that the various individual microcontacts that form a metainterface are independent. This means that the contact on one asperity is not affected by the presence of neighbouring microcontacts. For asperities placed at the surface of the same elastic object, this assumption is a priori wrong, because any local indentation causes a long-ranged normal downward displacement which is everywhere finite. In particular, when the highest asperity gets into contact with the indenting plane, all other asperities, not yet in contact, are displaced downward. Their heights, which had been carefully designed to produce a desired behaviour law, are actually not the ones expected when assuming independence, and the metainterface behaviour law is presumably modified.

When the deformable solid is a linearly elastic half-space, and when a Hertz contact of radius $a$ is created, the induced downward displacement of the rest of the plane is \cite{johnson1987contact}:
\begin{equation}
U_z(r)=-\frac{a^2}{\pi R}\left[\left(2-\frac{r^2}{a^2}\right)sin^{-1}\left(\frac{a}{r}\right)+\frac{r}{a}\sqrt{1-\frac{a^2}{r^2}}\right],\label{eq:interactions}
\end{equation}
\noindent where $r>a$ is the distance to the contact center and the vertical displacement of the surface of the half-space, $U_z$, is taken positive outward the solid. $U_z$ decays as $\sim$$1/r$, so that elastic interactions tend to vanish when the microcontacts are placed increasingly farther. But the practical question for metainterfaces is: how far apart does one need to place asperities in order to neglect elastic interactions? A fact is that, if microcontacts are not independent, then the placement of the 64 asperities with different heights on the nodes of the $8\times8$ square lattice becomes relevant. Different placements will presumably yield different behaviours due to different neighbourhoods for the microcontacts.

\subsection{Effect of asperity placement on the lattice}\label{sec:permutation}

In this section, we assess the role of the placement of asperities within the lattice that defines the surface texture, by considering select permutations of the asperities in the lists of heights of~\App{sec:appendixA}. Those permutations, which do not affect the heights distributions, are denoted as variants. All variants used in the figures of this article are reported in the tables of~\App{sec:appendixB}.

We first consider handmade, but patternless, shuffling of the asperities, for metainterfaces QL1, QL2, BL1 and BL2 (\REV{as an illustration, compare in \Fig{fig:QL2variants} the reference and variant 1 of QL2}). In all cases, such permutation produces negligible changes to the compression laws (as illustrated on~\Fig{fig:shuffledQL2} for the case of QL2, variant 1). This result suggests that patternless shuffling does not significantly impact the average neighbouring of the microcontacts.

\begin{figure}[ht!]
\centering
\begin{subfigure}[b]{0.49\linewidth}
    \includegraphics[width=\linewidth]{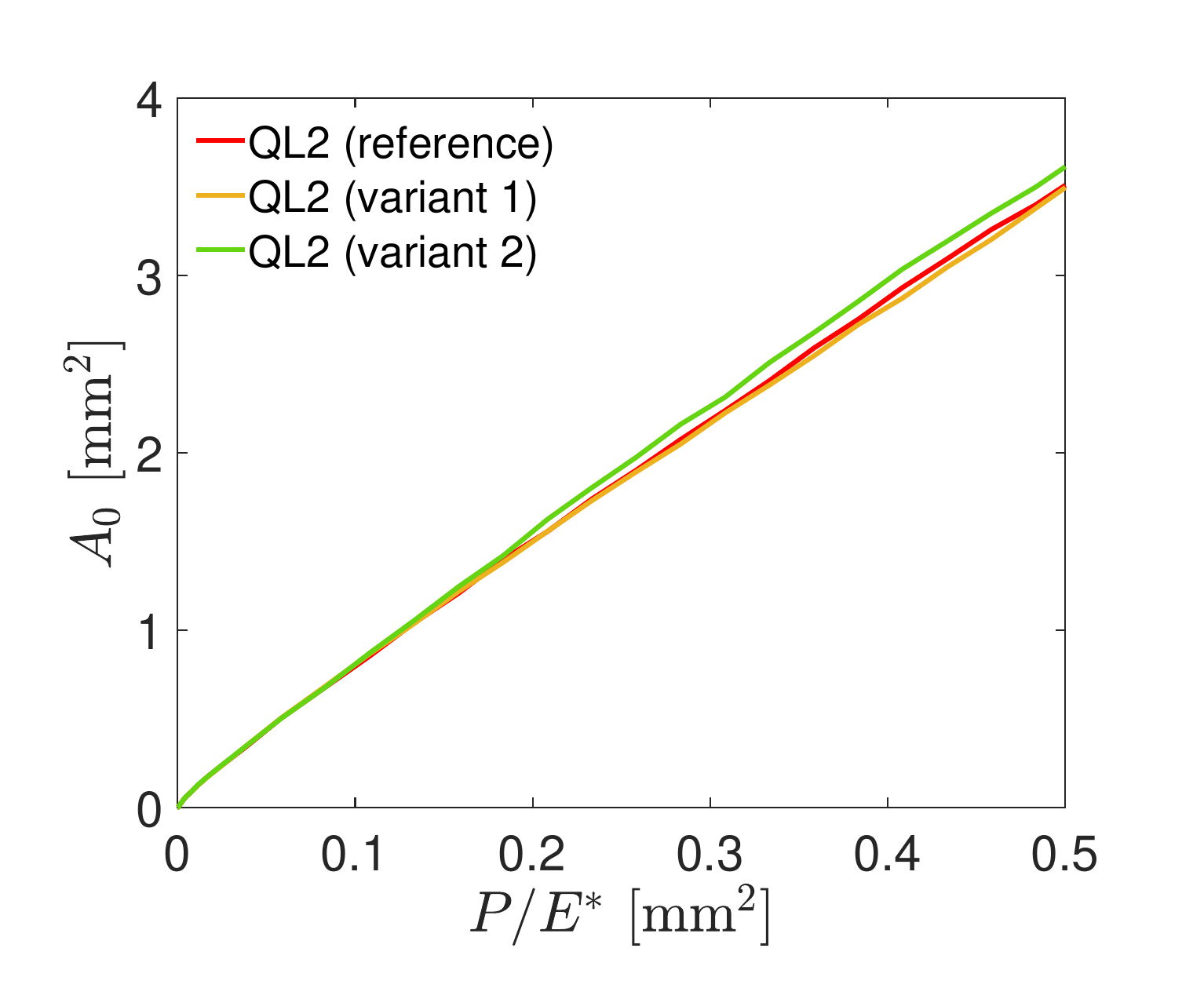}
    \caption{Quasilinear law 2 (QL2) with permuted asperity locations.}
    \label{fig:shuffledQL2}
\end{subfigure}
\begin{subfigure}[b]{0.49\linewidth}
    \includegraphics[width=\linewidth]{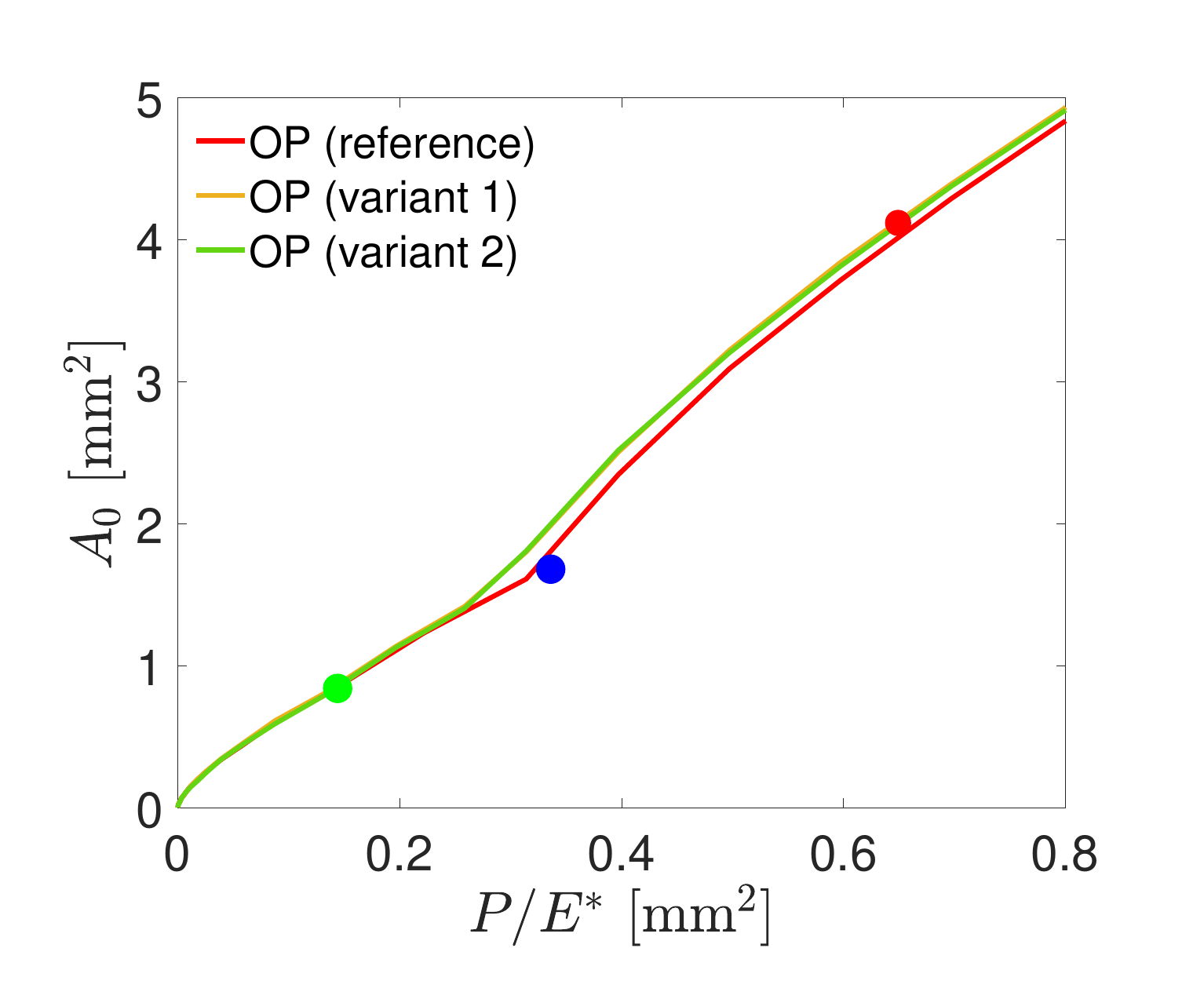}
    \caption{Operating points law (OP) with permuted asperity locations.}
   \label{fig:shuffledOP}
\end{subfigure}
\begin{subfigure}[b]{0.9\linewidth}
    \includegraphics[width=\linewidth]{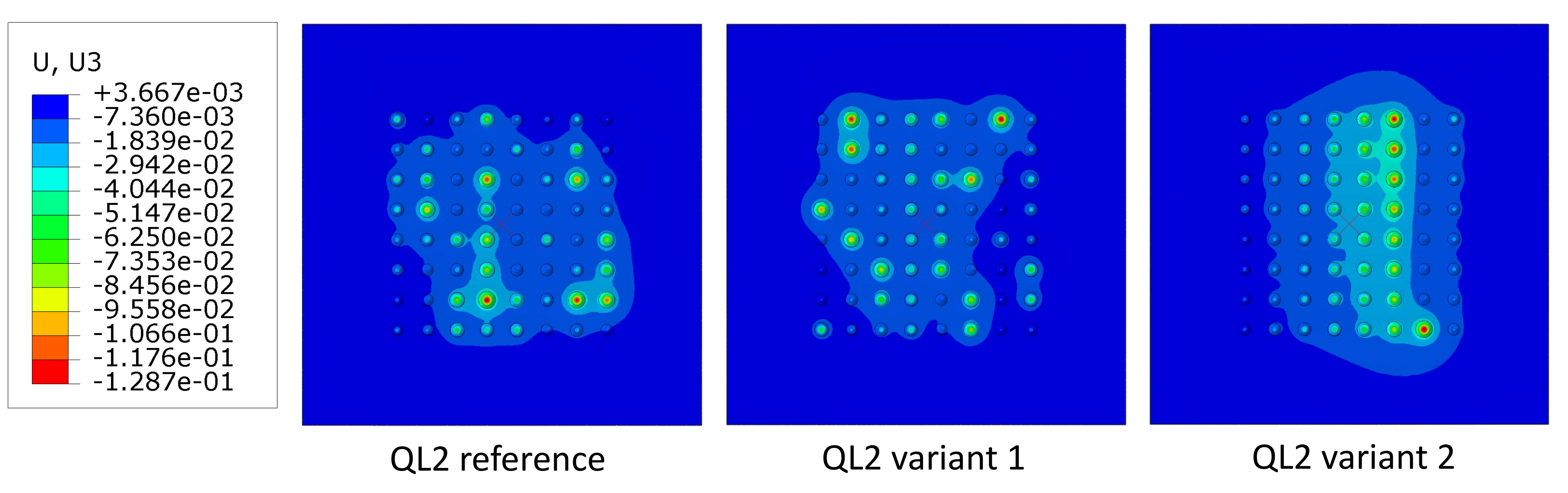}
    \caption{\REV{QL2 metainterface and its variants.}}
   \label{fig:QL2variants}
\end{subfigure}
\begin{subfigure}[b]{0.9\linewidth}
    \includegraphics[width=\linewidth]{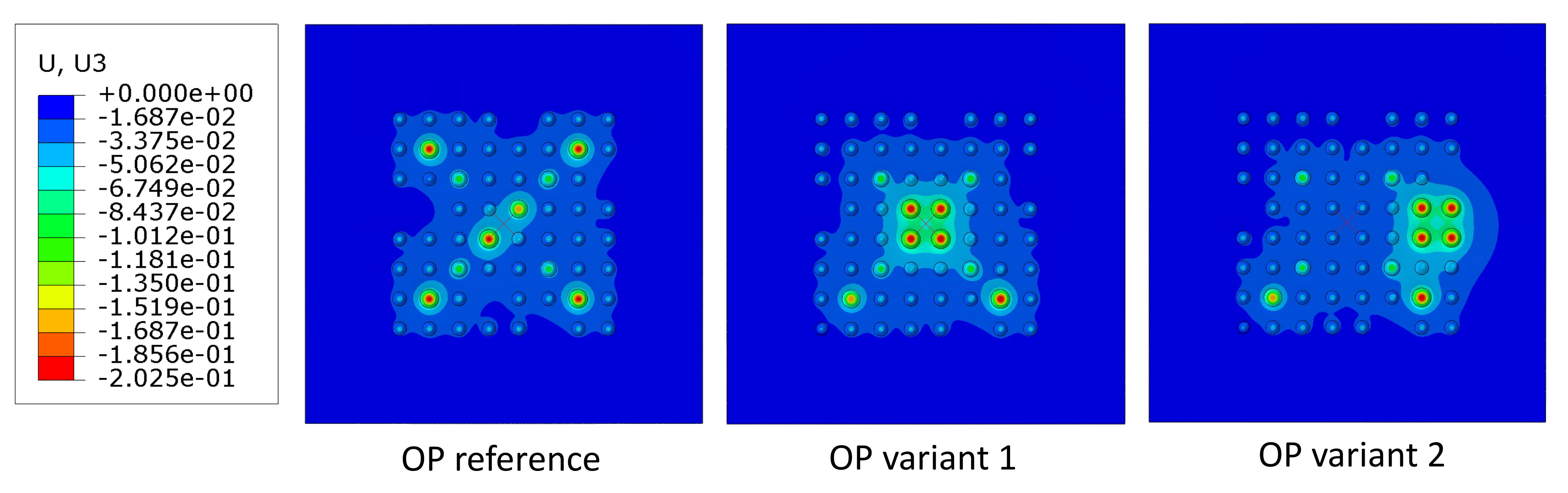}
    \caption{\REV{OP metainterface and its variants.}}
   \label{fig:OPvariants}
\end{subfigure}
\caption{Effect of permuting asperity locations on the compression behaviour. (a)--(b): Contact area $A_0$ as a function of the normal force, $P$, for metainterfaces QL2 (a) and OP (b). \REV{Reference metainterfaces are the ones }with height lists given in~\App{sec:appendixA}. \REV{Variants are based on permuted asperities with height lists given in}~\App{sec:appendixB} (see legend for the variant index). \REV{(c) (resp. (d): Fields of vertical displacement, $U_z$ (noted $U_3$ in the legends), for the metainterface QL2 (resp. OP) and its variants, for a normal force, $P/E^*$=0.5\,mm$^2$ (resp. 0.8\,mm$^2$).}
The unit of $U_z$ is in mm.} 
\label{fig:shuffled}
\end{figure}

To test this hypothesis, for QL2, we define variant 2 made of linear blocks of 4 asperities with identical or close heights. For instance, in this variant, the sixth column gathers 8 of the highest asperities (see \Fig{fig:QL2variants}). \Fig{fig:shuffledQL2} shows that this extreme variant corresponds to a limited, but clearly noticeable impact on the compression law. This observation suggests that elastic interactions may manifest when high asperities are placed on neighbouring nodes of the lattice.

This interpretation is further supported by complementary permutation tests performed on the OP metainterface. For this metainterface, the reference placement is not random: the highest asperities are located along the diagonals of the square lattice (\REV{see OP reference in \Fig{fig:OPvariants}}). We first test a permutation that places four of the largest asperities at the center of the lattice (\REV{ see OP variant 1 in \Fig{fig:OPvariants}}). Doing so, a non-negligible change is observed on the compression behaviour, with larger contact areas than for the reference, beyond $P/E^*$$\simeq$0.3\,mm$^2$. As shown on \Fig{fig:OPvariants}, this configuration involves larger regions with increased downward vertical displacement. For instance, the region between the four central asperities in variant 1 has displacement about 100\,$\mu$m, with no equivalent in the reference interface, for the same normal force. The compression behaviour is found identical when using variant 2, where the four central asperities are shifted on the side of the lattice (see \REV{OP variant 2 in \Fig{fig:OPvariants}}). This observation confirms that, for these dimensions of the elastic base, the change in the behaviour compared to the reference placement is mainly due to the patterns made of four neighbouring high asperities, and not to the precise location of this pattern within the lattice.

The fact that, in \Fig{fig:shuffledOP}, the second (blue) operating point is missed, relates to the constitution of OP's height distribution in three main levels (one for each operating point). Such peculiar distribution enhances two distinct effects related to elastic interactions. First, as previously shown, e.g., in~\cite{ciavarella2008inclusion}, interactions reduce the stiffness of an assembly of microcontacts. Consistently, we have observed that the presence of a cluster of high asperities in variants 1 and 2 requires a larger indentation $\delta$ to reach a given $P/E^*$. This effect implies that the group of asperities forming the next height level comes into contact for a smaller normal force than for the reference configuration where high asperities are farther apart. This consequence is clearly visible as a shift to the left of the main kink in the compression curve (at $P/E^*$$\simeq$0.3\,mm$^2$) when passing from the reference configuration (red curve) to the variants (other curves). The second effect of interactions is that a group of asperities initially at the same level have distributed heights when the cluster is indented. The asperities that are closer to the cluster are brought further down and get into contact at a later stage, leading to a noticeably smaller area in the branch where $P/E^*$$>$0.3\,mm$^2$.

We emphasize that the two above-mentioned effects of interactions (increased area due to a larger indentation for a given $P$ and delayed contact due to existing closeby contacts) have opposing consequences on the contact area. In a given metainterface, they are generally entangled, so that their overall result on the compression curve are very difficult to predict.

\subsection{Effect of the interdistance between asperities}

For the metainterface dimensions used in experiments, we have seen that a non-negligible effect of the placement of asperities can only be seen for very specific configurations, where large asperities are placed next to each other. But, as seen in Eq.~\ref{eq:interactions}, the elastic displacement around a microcontact increases as the distance from the microcontact decreases. One thus expects that asperities that are closer to each other will experience a larger influence from their neighbours. To investigate this effect, we modify the interdistance between asperities by changing the pitch $d$ of the square lattice.

In the reference metainterfaces of \Fig{fig:DirectSimu}, the pitch was $d$=1.5\,mm. Here, we investigate the effect of varying $d$ between 1 and 2.5\,mm (for $d$ below 1\,mm, the diameter of the base of the spherical caps corresponding to  the highest asperities would be larger than $d$, a situation that we do not consider here), while keeping the size of the elastic base unchanged. Our main result (not shown) is that, for all metainterfaces and their variants, the effect of varying $d$ is very limited. The compression curves with $d$ varying between 1 and 2.5\,mm are actually almost indistinguishable in all cases except QL2 variant 2 and OP variants 1 and 2. Note that those three metainterfaces correspond precisely to the asperity permutations that produce noticeable discrepancies with the reference configuration (see~\Fig{fig:shuffledQL2} and \Fig{fig:shuffledOP}). Such correspondence strongly suggests that the clusters of high asperities that are present in those three metainterfaces are the parts that are the most sensitive to the interdistance $d$ between asperities.

\Fig{fig:interdistance}a shows the evolution of the compression law of metainterface QL2 variant 2, when $d$ is either decreased to 1\,mm or increased to 2.5\,mm from the reference interasperity distance $d$=1.5\,mm. As for the permutations in~\Fig{fig:shuffledQL2}, the curves exhibit discrepancies only above $P/E^*$$\simeq$0.2\,mm$^2$. In this range, a slight trend is observed, where the curves get lower as $d$ increases, and tend toward that of the reference asperity placement. This convergence is not unexpected because a negligible effect of interactions may be obtained either by increasing $d$ or by using a patternless asperity placement, as in the reference configuration (see \Sec{sec:permutation}).

\begin{figure}[ht!]
\centering
\begin{subfigure}[b]{0.49\linewidth}
    \includegraphics[width=\linewidth]{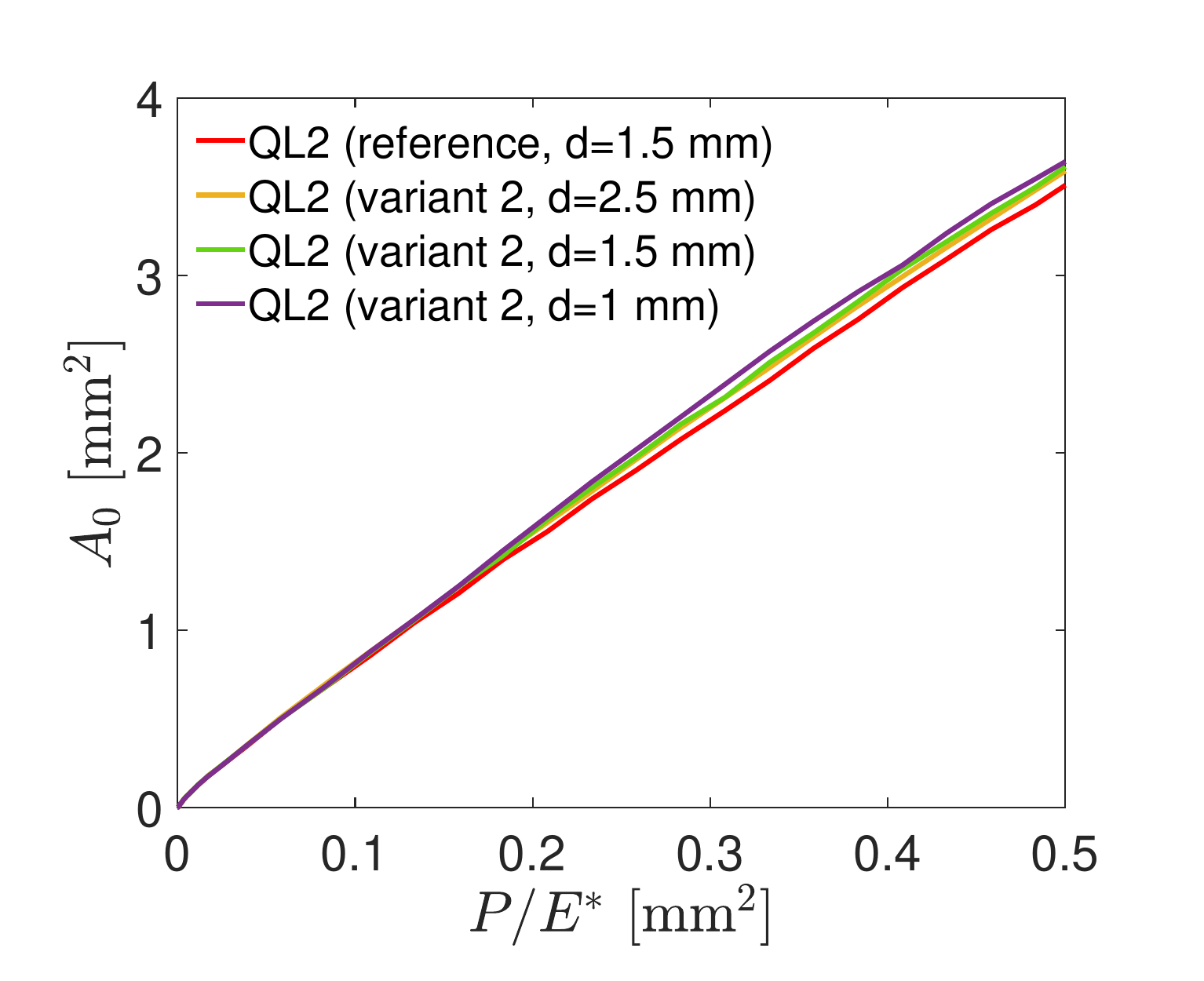}
    \caption{Quasilinear law 2 (QL2).}
    \label{}
\end{subfigure}
\begin{subfigure}[b]{0.49\linewidth}
    \includegraphics[width=\linewidth]{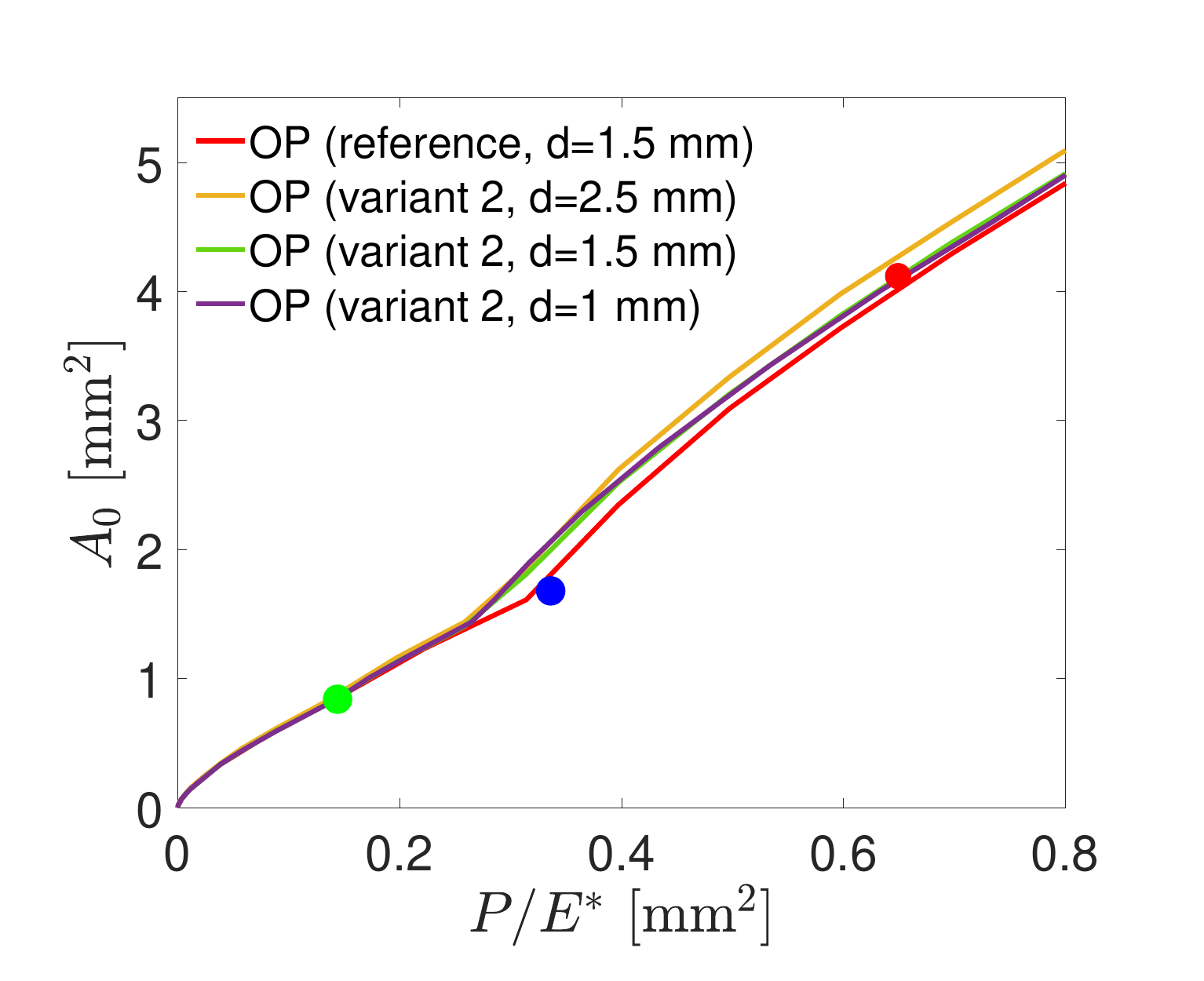}
    \caption{Operating points law (OP).}
   \label{}
\end{subfigure}
\caption{Effect of varying the interdistance between asperities, $d$, on the metainterfaces where this effect is found maximum. Contact area, $A_0$, versus $P/E^*$ for (a) metainterface QL2, variant 2 and (b) metainterface OP, variant 2. Red lines: reference asperity placement (same data as red lines in~\Fig{fig:shuffled}a and b). Purple, green and yellow lines correspond, for variants 2, to $d$=1, 1.5 and 2.5\,mm, respectively (see legend). The green lines are the same as the green lines in~\Fig{fig:shuffled}a and b.} 
\label{fig:interdistance}
\end{figure}

 \Fig{fig:interdistance}b shows similar results, but for metainterface OP, variant 2 (the results for variant 1 are indistinguishable from those of variant 2 and are thus not shown). The curve obtained for $d$=1\,mm (purple) is almost identical to the one for the reference interdistance 1.5\,mm (green), the shape of which (see \Fig{fig:shuffled}) has already been interpreted in detail in \Sec{sec:permutation}. 
 For $d$=2.5\,mm, where interactions are weak, almost all asperities in the last height level touch simultaneously, leading to a larger area than for smaller $d$, all along the last branch (yellow curve in~\Fig{fig:interdistance}b). Notice that this branch has the same shape (although shifted) as in the reference case (red line), most likely because interactions are also weak in this latter case due to the absence of cluster, so that all asperities of the last height level also touch rather simultaneously. This same branch may also be influenced by the proximity of the asperities from the border of the base, as discussed in \Sec{sec:lateral}.

\subsection{Rationale}

All the above observations may be rationalized based on Eq.~\ref{eq:interactions}. This equation suggests that the relevant dimensionless parameter that controls elastic interactions on microcontact $i$ is $a_i/r_{ij}$, where $a_i$ is the microcontact radius and $r_{ij}$ is the distance between microcontacts $i$ and $j$. Assuming, for the argument's sake, that interactions become non-negligible when $a_i/r_{ij}$ reaches a threshold value, this threshold is progressively approached when the compression on the metainterface increases, through two simultaneous mechanisms. First, $a_i$ increases for all existing microcontacts. Second, new microcontacts are created, which decreases the distance $r_{ij}$ between existing microcontacts.

For small normal forces, $a_i$ is small and $r_{ij}$ is large (because only few microcontacts exist), so that $a_i/r_{ij}$ is always below the threshold. In this regime, interactions remain negligible, explaining why the beginning of the compression curves is independent, for a given list of heights, of the particular placement of the asperities and of the interasperity distance $d$. For large normal forces, e.g. when $P$ reaches a maximum experimental value, $P_{max}$, a maximum contact radius, $a_{max}$, is obtained in the metainterface. Noting that the minimum value of $r_{ij}$ is $d$, if the interaction threshold is larger than $a_{max}/d$, then the elastic coupling cannot affect the behaviour in the observed range [0 ; $P_{max}$] and the metainterface responds as if microcontacts were independent.

The effect of interactions becomes all the more visible as the threshold is reached at a smaller normal force. An efficient way of maximizing $a_i/r_{i,j}$ is to place large asperities at adjacent nodes of the lattice. This configuration, which both minimizes $r_{i,j}$ to $d$ and maximizes $a_i$, precisely corresponds to the clusters of high asperities that have been created in variants 1 and 2 of metainterface OP and in variant 2 of metainterface QL2. This explains why those configurations are the ones exhibiting the largest impacts of permutations and change of inter-asperity distance. Other asperity placements presumably keep $a_i/r_{ij}$ to a value large enough to prevent the threshold to be reached within the explored range of compression.


\section{Finite size effects}\label{sec:FSE}

In the previous section, we have seen that contact on one asperity can influence the contact on all other asperities due to long-ranged elastic deformations. These long-ranged deformations also make it such that the contact state is a priori sensitive to the finite size of the elastic parallelepipedic base on which the asperities are placed. Indeed, the boundary conditions at the fixed bottom and stress-free lateral sides of the base strongly differ from the half-space assumption underlying Eqs.~\ref{Eq:A} and~\ref{Eq:P}. In~\cite{aymard2024designing}, the sample thickness $H$ was more than 14 times larger than the asperity radius, $R$, and the total surface of the asperity-bearing face was about 4 times larger than the surface of the square lattice of asperities. Those empirically-chosen values were expected to ensure that the half-space assumption is sufficiently good. Here, we test the validity of this expectation by running FE simulations with varying lateral and vertical dimensions of the base.

\subsection{Effect of the finite lateral size of the base}\label{sec:lateral}

As visible in~\Fig{fig:model}, for the reference inter-asperity distance $d$=1.5\,mm, the textured square is 12\,mm\,$\times$\,12\,mm, so that all asperities are located at a distance larger than $w$=4\,mm from the border of the sample. Note that, when using $d$=2.5\,mm, as sometimes done in \Sec{sec:interactions} (see, e.g., yellow lines in \Fig{fig:interdistance}), $w$ reduces to 0.

\begin{figure}[ht!]
    \centering
    \includegraphics[width=0.725\linewidth]{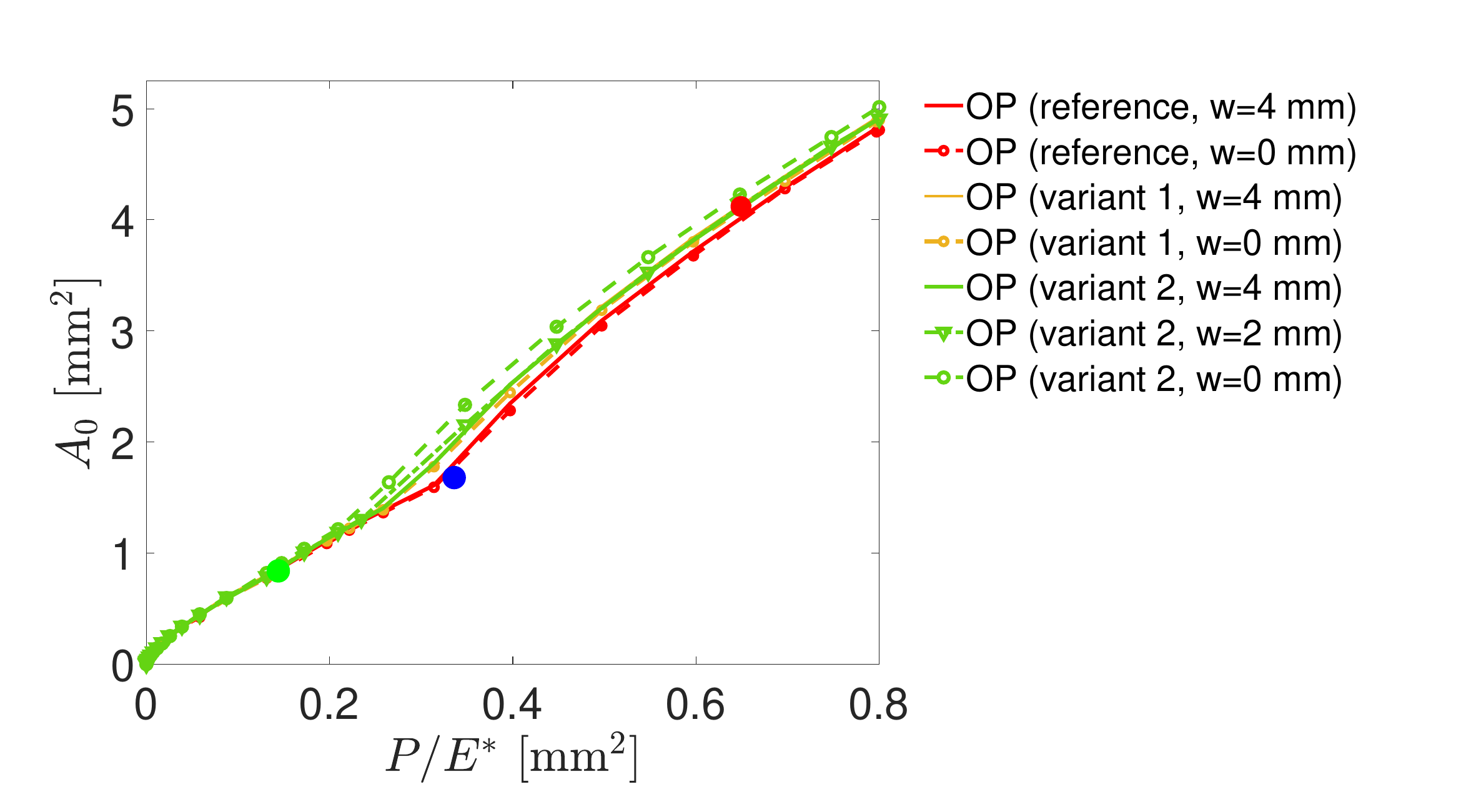}
    \caption{Effect of varying the lateral size of the elastic base at constant size of the square lattice ($d$=1.5\,mm.), for metainterface OP. $A_0$ vs $P/E^*$ for the reference asperity placement (red curves) and variants 1 (yellow) and 2 (green). Solid lines are for the reference distance of the lattice to the base's border, $w$=4\,mm. Open disks and triangles are for $w$=0 and 2\,mm, respectively. Solid disks are the three target operating points.}
    \label{fig:lateral2}
\end{figure}

In this section, we modify the value of $w$ for some metainterfaces already used in \Sec{sec:interactions}, while keeping $D$=20\,mm. Our main observation is that reducing $w$ down to 0 brings negligible changes to the compression behaviour of most metainterfaces, including all reference ones. This observation indicates that the combination of $d$=1.5\,mm and a sample of lateral size 20\,mm is adequate to use the design strategy of~\cite{aymard2024designing}, except when high asperities are placed on the sides of the square lattice (e.g. for QL variant 2 and OP variant 2). Such exception is illustrated, on the example of metainterface OP, in~\Fig{fig:lateral2}. For the reference asperity placement, the curves for $w$=4 or 0\,mm are indistinguishable (compare red lines in \Fig{fig:lateral2}). 
For variant 1 with a cluster of high asperities placed at the center of the lattice, reducing $w$ down to 0 has a detectable but very limited effect (compare yellow lines in \Fig{fig:lateral2}). Instead, for variant 2, where the same cluster is now placed on a side of the lattice, a significant effect is observed on the last branch of the curve when passing from $w$=2 to 0\,mm (compare green triangles and disks in \Fig{fig:lateral2}): the branch starts at a reduced $P/E^*$ and the area is larger all along the branch. Such effect is further amplified when using smaller values of $d$ (e.g., 1\,mm instead of 1.5\,mm, not shown).

The observed effect can be understood as follows. As shown in \Fig{fig:monoH}, a microcontact close to the sample's border (green line) has the same $a_0(p)$ evolution (\Fig{fig:monoH}a) but experiences an effectively smaller stiffness (\Fig{fig:monoH}b) than a central one (blue line), due to the absence of elastic medium on one side. Thus, the indentation needed to reach a given $P$ is larger, so that the next level of asperities gets into contact at a smaller $P/E^*$, explaining both an earlier onset for the branch, and a larger total area in this branch.

\begin{figure}[ht!]
\centering
\begin{subfigure}[b]{0.49\linewidth}
    \includegraphics[width=\linewidth]{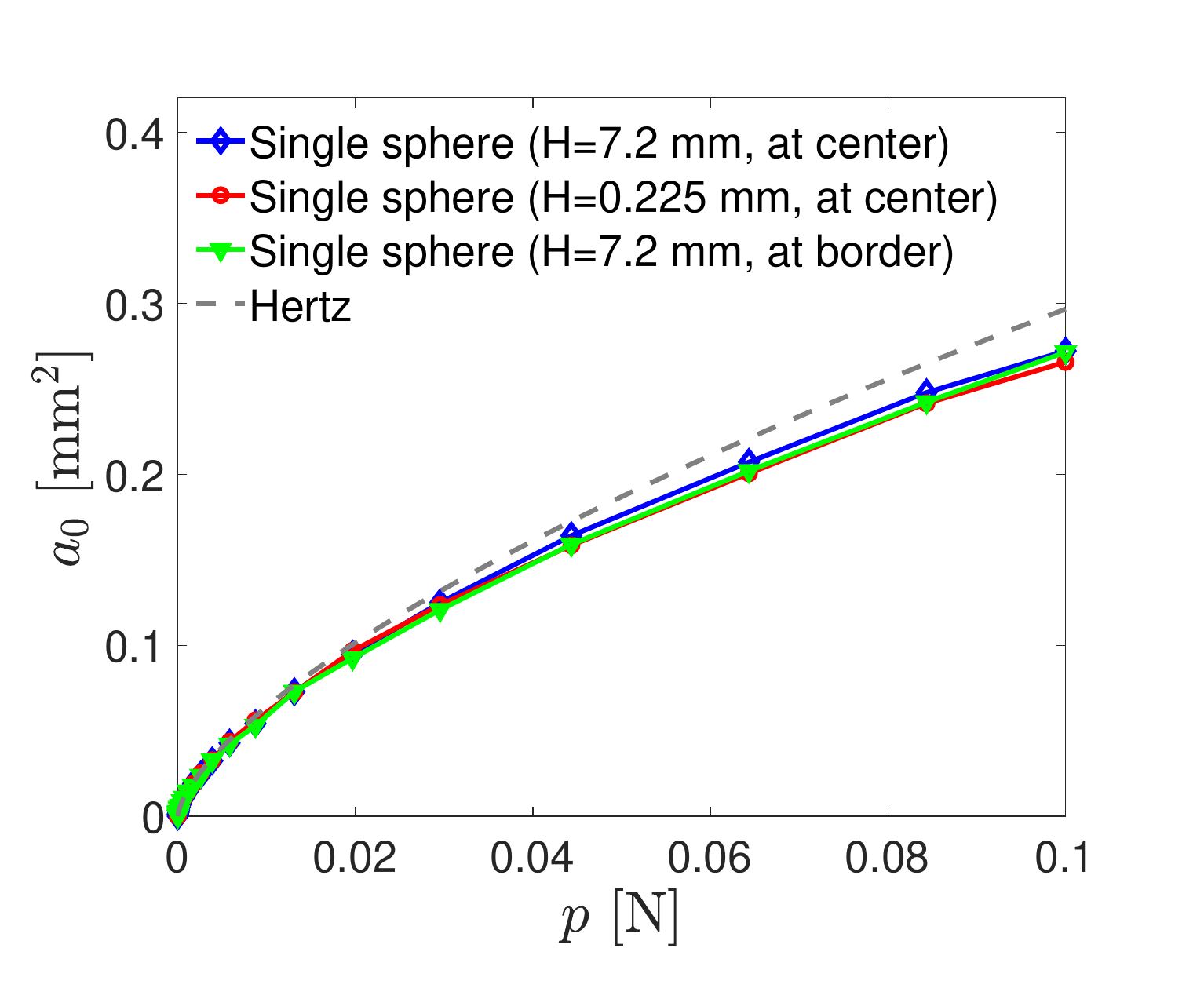}
    \caption{$a_0$ as a function of $p$.}
    \label{}
\end{subfigure}
\begin{subfigure}[b]{0.49\linewidth}
    \includegraphics[width=\linewidth]{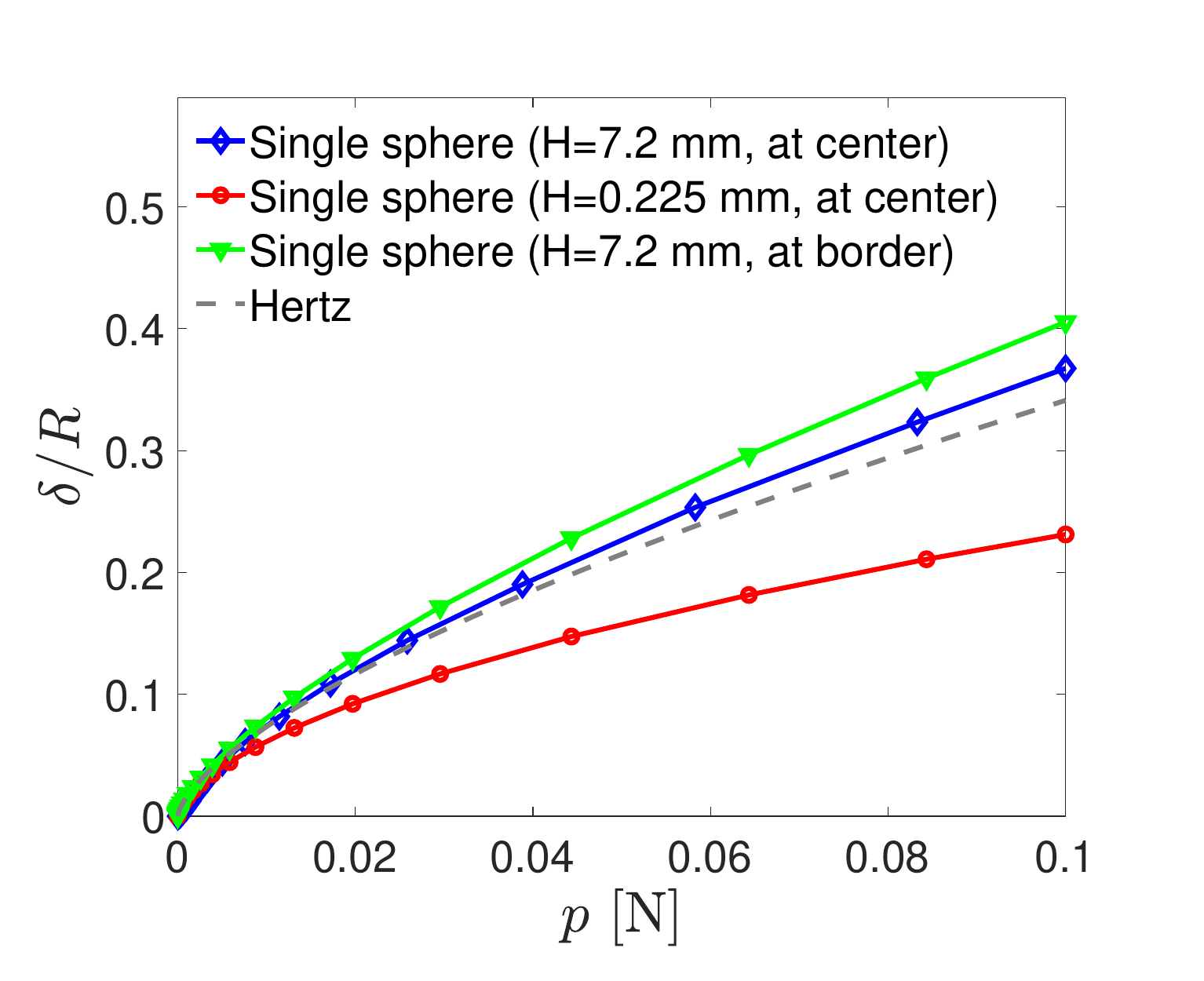}
    \caption{$\delta\REV{/R}$ as a function of $p$.}
   \label{}
\end{subfigure}
\caption{Effect of the finite size of the sample on the compression behaviour of a single asperity (asperity height 270\,$\mu$m). (a) $a_0$ vs $p$. (b) $\delta\REV{/R}$ vs $p$. Dashed line: Hertzian behaviour. Blue solid line \REV{and diamonds}: simulated behaviour for the reference thickness $H$=7.2\,mm and for an asperity at the center of the sample (same data as solid blue line \REV{and diamonds} in \Fig{fig:onesphere}). Green solid line and triangles: Same simulation but for an asperity at the border of the sample (asperity center at 0.75\,mm of the border, as in the case $d$=1.5\,mm, $w$=0\,mm). Red solid line and disks: same simulation as blue line, but with $H$=0.225\,mm.} 
\label{fig:monoH}
\end{figure}

\subsection{Effect of the finite thickness of the base}\label{sec:thickness}

\Fig{fig:thickness} shows the evolution of the curve $A_0(P)$ for metainterface OP, as the sample's thickness $H$ decreases from its reference value of 7.2\,mm down to 0.225\,mm, while keeping $D$=20\,mm, $d$=1.5\,mm and $w$=4\,mm. Contrary to what is observed when changing $w$ (see \Fig{fig:lateral2}), significant effects can be seen already on the reference asperity placement. Dividing $H$ by 2 gives indistinguishable behaviour, showing that the reference thickness 7.2\,mm was amply sufficient to apply the design strategy described in~\cite{aymard2024designing}. Non-negligible effects appear for thicknesses below about 1\,mm, and tend to shift the onset of the curve's last branch to larger $P/E^*$. For $H$=0.225\,mm, the onset occurs at $P/E^*$ as large as about twice its reference value, so that the third operating point (red disk) is far from being approached and a design based on a half-space assumption would fail.

\begin{figure}[ht!]
    \centering
    \includegraphics[width=0.5\linewidth]{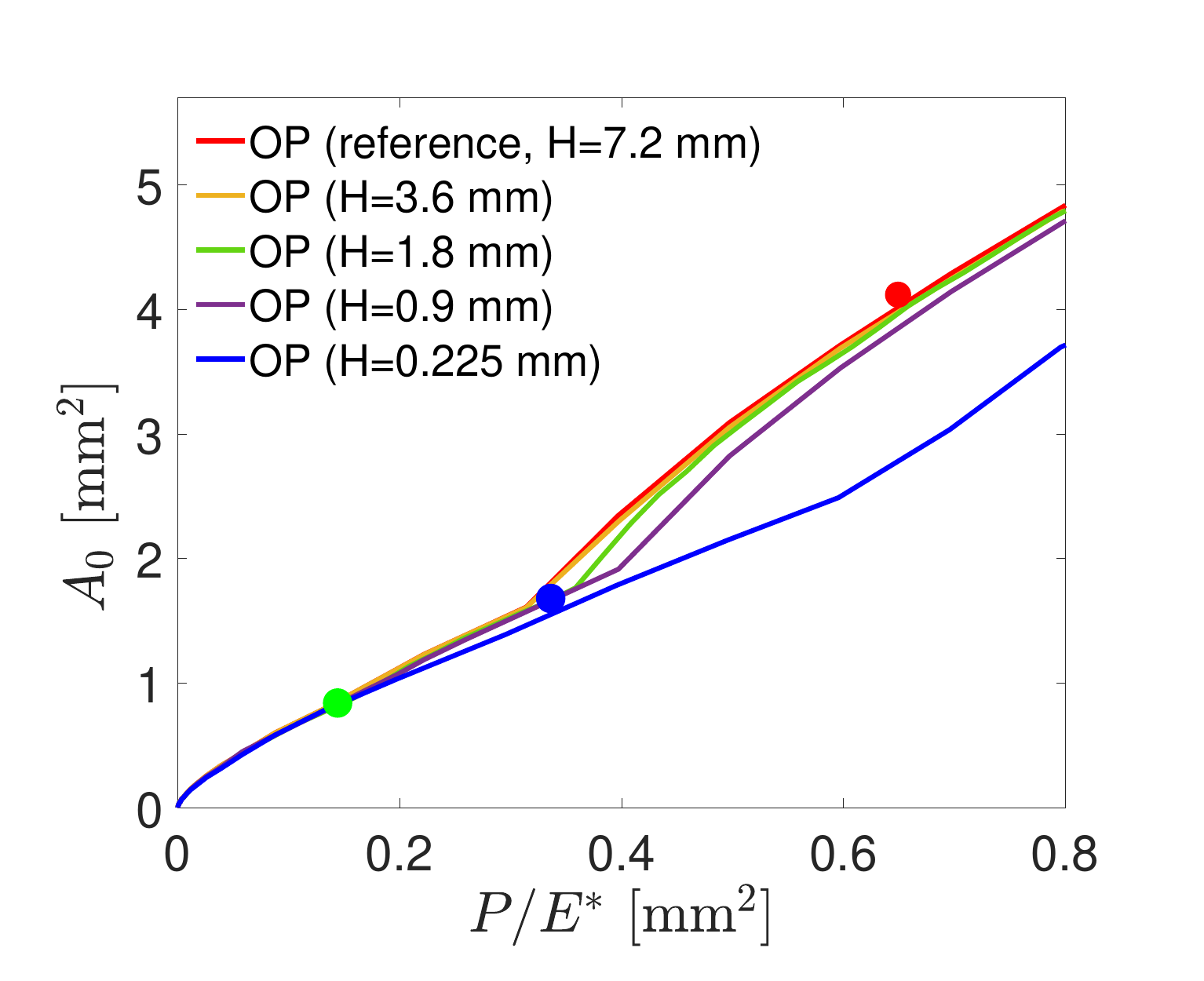}
    \caption{Effect of varying the sample's thickness, for metainterface OP with the reference asperity placement. $A_0$ vs $P/E^*$ for various $H$, from the reference value $H$=7.2\,mm down to 0.225\,mm (see legend). Solids disks are the three target operating points.}
    \label{fig:thickness}
\end{figure}

The explanation of this behaviour is qualitatively similar to that used to interpret the effect of the reduction of $w$ in \Sec{sec:lateral}, but with opposite causes and results. Reducing $H$ affects significantly the $\delta(p)$ evolution (see \Fig{fig:monoH}b, compare blue and red curves). As discussed, e.g., in~\cite{dimitriadis2002determination}, a thinner base below the asperity corresponds to a stiffer contact where a smaller indentation is required to reach a given $p$. In the OP metainterface, this individual behaviour is such that the indentation necessary to create contact with the group of asperities responsible for the last branch corresponds to a much larger $P$ than for the reference thickness. This is why the third branch of the blue curve onsets much later and the curve remains much smaller than for the reference one (red curve).

\section{Discussion}\label{sec:discussion}

\REV{To assess whether the predicted effects of finite size and microcontact interactions can be detected experimentally, we can compare two curves: First, the absolute value of the computed difference in contact area between the compressive laws from a reference configuration and a given variant, denoted $\vert \mathrm{\Delta}  A_0(P) \vert$ ; second, the evolution with $P/E^*$ of the experimental error bars on $A_0$ shown in \Fig{fig:DirectSimu}. Representative results of such comparisons are shown in \Fig{fig:deltaA}, for metainterfaces QL2 and OP (panels a and b, respectively). A predicted difference between configurations is expected to be experimentally detectable wherever $\vert \mathrm{\Delta} A_0(P)\vert$ (solid lines) is larger than the experimental uncertainty (markers and dashed line).}


\textcolor{red}{\begin{figure}[ht!]
\centering
\begin{subfigure}[b]{0.49\linewidth}
    \includegraphics[width=\linewidth]{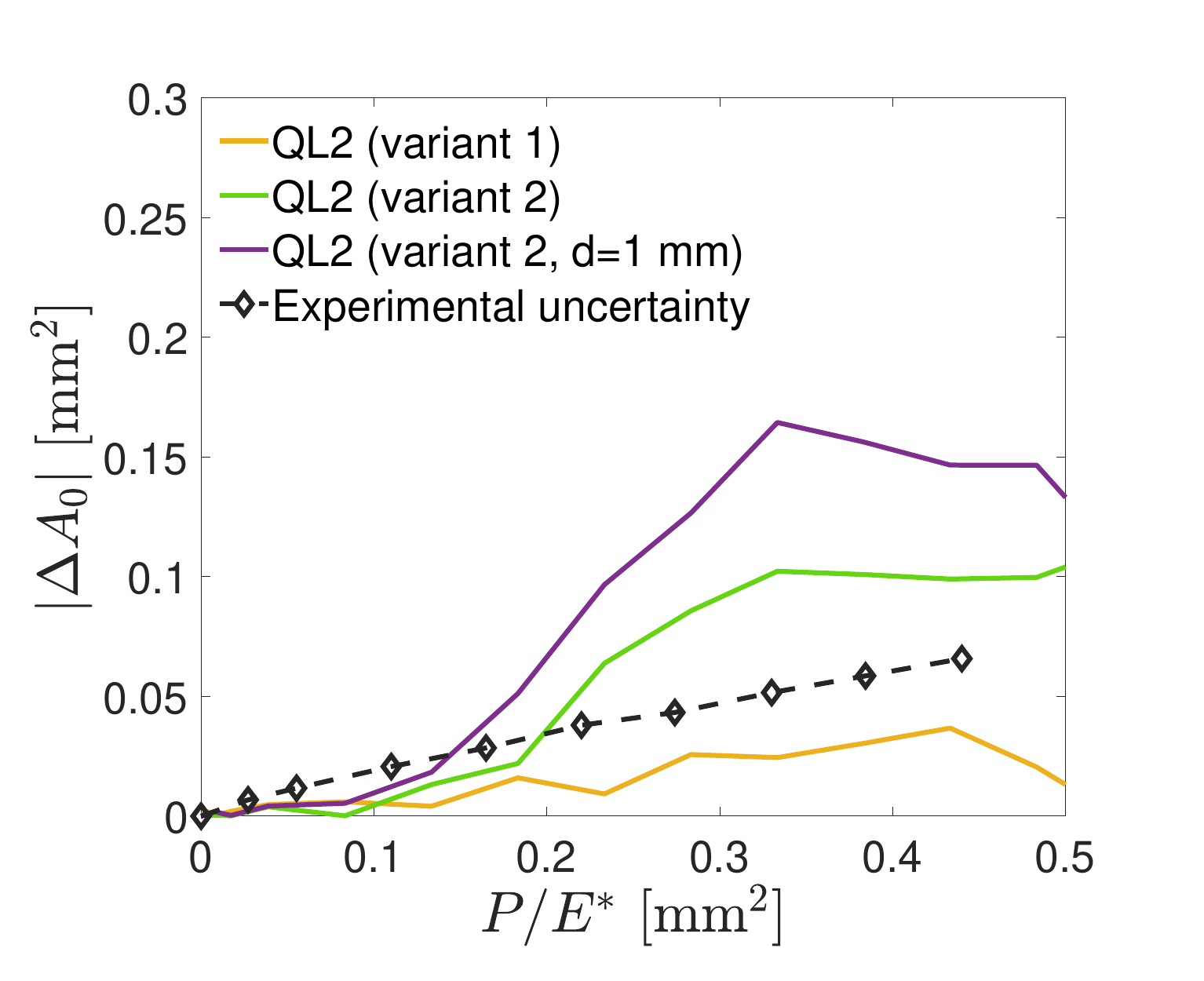}
    \caption{Quasilinear law 2.}
    \label{fig:deltaA_QL2}
\end{subfigure}
\begin{subfigure}[b]{0.49\linewidth}
    \includegraphics[width=\linewidth]{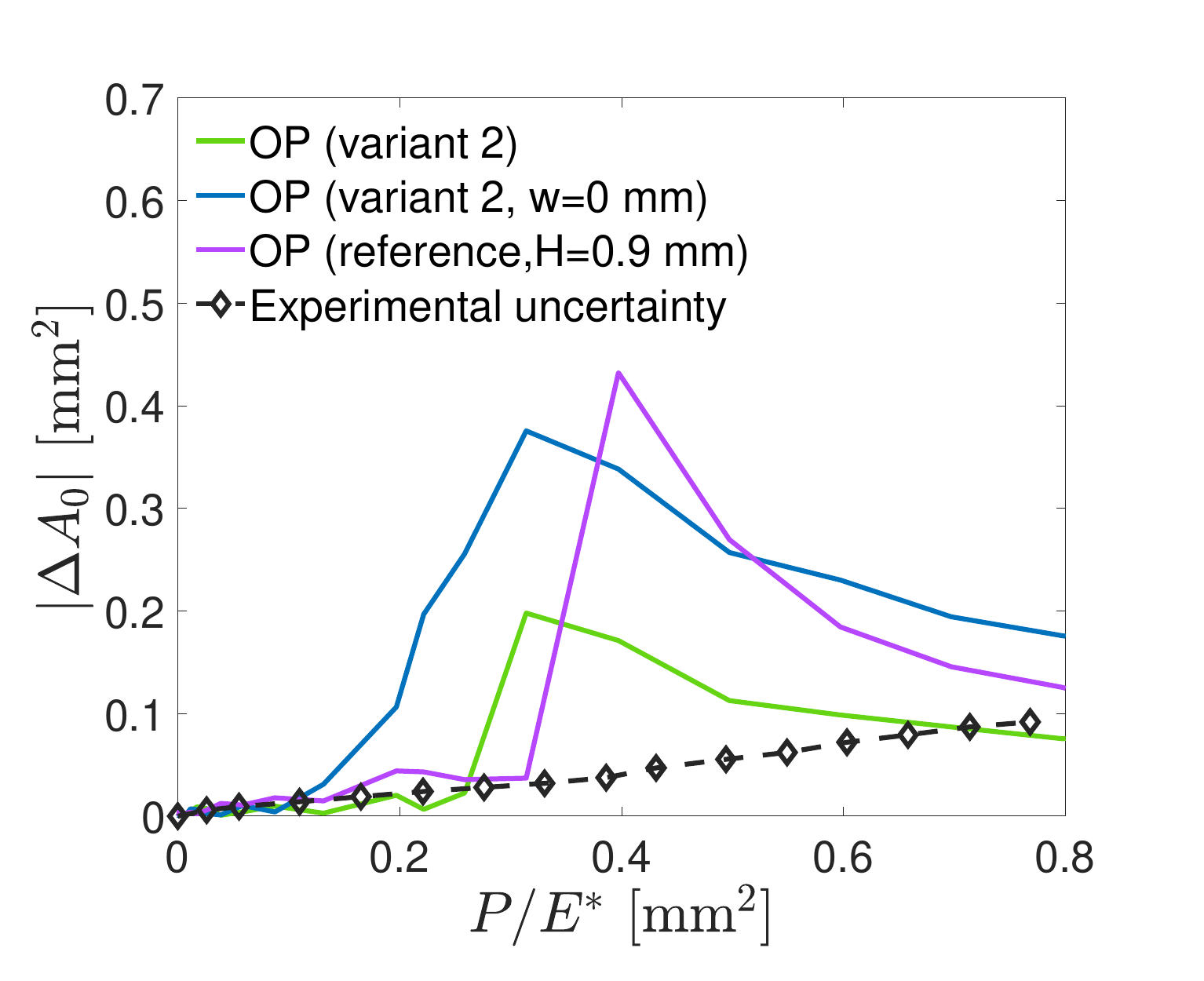}
    \caption{Operating points law.}
   \label{fig:deltaA_OP}
\end{subfigure}
\caption{\REV{Solid lines: Evolutions of the difference in contact area between reference configurations and varied ones, $\vert \mathrm{\Delta} A_0(P)\vert$, as a function of $P/E^*$, for (a) the QL2 and (b) the OP metainterface. Dashed lines and markers: experimental uncertainty on the contact area, taken from panels (b) and (e) of \Fig{fig:DirectSimu}.}} 
\label{fig:deltaA}
\end{figure}
}

\REV{\Fig{fig:deltaA_QL2} illutrates that variants based on patternless shuffling of asperities, like variant 1 of QL2, are essentially undistinguishable from the reference configuration, because $\vert \mathrm{\Delta} A_0(P)\vert$ (yellow line) is everywhere below the experimental uncertainty (dashed line). Instead, for variant 2, where clusters of high asperities exist, their effect is detectable at sufficiently large normal load (green line, for $P/E^*$ beyond 0.2\,mm$^2$). These interaction effects are amplified when the inter-asperity distance $d$ is reduced (purple line). A similar analysis can be done on the metainterface OP (\Fig{fig:deltaA_OP}). The already detectable effect of clustering in variant 2 (green line, especially close to the second operating point) becomes major when the sample border is very close to the asperities ($w$=0 mm, blue line) or when the sample thickness decreases below 1\,mm (purple line). Overall, most of the effects presented in \Sec{sec:interactions} and \Sec{sec:FSE} are amenable to experimental testing with existing measurement methods.}

\MOV{We emphasize that the \REV{identified} sensitivities of metainterfaces to geometrical parameters that experimentalists need to choose (but that are not explicitly considered in the design strategy) depend on the target behaviour law. Laws that target a sudden and large increase of the slope of the $A_0(P)$ curve (like for the metainterface OP) are particularly sensitive. An abrupt slope change requires many asperities to get into contact simultaneously, which corresponds to a large number of asperities having very similar heights. Such a peak in the height distribution makes the metainterface little robust to interactions. First, the reduced stiffness of clustered asperities compared to the same number of sparse asperities makes it such that the height peak is reached sooner than expected and that the change of slope occurs at a smaller normal force than targeted. Second, the asperities that are closer to the cluster will get into contact later than farther ones, so that the slope change is smoothed out and its abruptness is lost. More geneally, interactions affect the activation sequence of the asperities, which is all the less robust as they are designed to get into contact in a narrower range of indentation.}

\section{Conclusion}\label{sec:conclusions}

Using full 3D finite element simulations of a finite-sized elastic base whose upper surface is decorated by spherical caps with prescribed heights and positions, we have assessed the main assumptions underlying the design strategy for frictional metainterfaces introduced in~\cite{aymard2024designing}. We have first demonstrated that the geometrical, asperity-based strategy to obtain specified contact and friction laws is fundamentally valid: for all available cases from the literature, the performed simulations, although relaxing most assumptions made in the design approach, agree quantitatively with the design predictions, without the need of any adjustable parameter. This agreement shows that the assumptions underlying the adopted Hertz contact model (including linear elasticity instead of hyperleasticity, and parabolic asperity shape rather than spherical) have only limited influences on the macroscopic metainterface behaviour. Note that our simulations match equally well the experimental measurements from~\cite{aymard2024designing}.

\REV{In this context, the main added value of our FE model has been to enable the exploration of the limits of the design strategy, and in particular} the potential impact of geometrical changes brought to the metainterfaces, not reported experimentally. To investigate the possible role of elastic interactions between microcontacts, we have permuted the asperity locations to change their neighbourhoods and we have changed the interdistance between asperities, $d$. To examine the influence of the finite size of the asperity-bearing elastic base, we have reduced either the base's thickness, $H$, or the distance between the borders of the lattice and of the base, $w$. In the explored ranges, most of those changes only bring negligible changes to the compression behaviour of the metainterfaces. Only in extreme conditions have we been able to trigger non-negligible deviations. Those conditions include the creation of clusters of high asperities or the presence of high asperities in the immediate vicinity of the border of the base. Such cases can easily be avoided in practice, by using base dimensions that ensure a distance to the border larger than about four times the asperity radius ($w\gtrsim 4R$ ; in our simulations where $R$$\simeq$0.5\,mm, $w$=2\,mm was always sufficient), and by distributing the asperities randomly among the nodes of their lattice.

In contrast, the base's thickness can have a large impact on the metainterface behaviour if it goes below about twice the asperity radius (in our case, when $H$ is smaller than about 1\,mm). We thus recommend to keep the thickness $H$ such that $H \gtrsim 10 R$ to avoid any finite-thickness effect.

\REV{We emphasize that the above-mentioned recommendations strictly apply to the compressive behaviour of metainterfaces. To conclude about the friction behaviour, an extension of this work to tangential loading and finite friction is required, likely involving  advanced solvers for frictional contacts~\cite{ladeveze2002multiscale,giacoma2015toward,mergel2021contact,zeka2024benefits,zeka2025control}. Indeed, long-ranged elasticity combined with specific neighbourings will likely also affect the shear and friction behaviours of individual asperities, as already seen for instance in the context of wear~\cite{pham2021adhesive}}.

Besides \REV{formulating} recommendations for a safe usage of the metainterface design strategy of~\cite{aymard2024designing}, our results point to a clear avenue for improving and extending the strategy. All the finite-size and interaction effects that we have highlighted can actually be turned into additional design levers to identify new and/or more robust surface arrangements for finely-tuned \REV{compression} behaviours.  \REV{While the precise identification of such new designs and the original behaviours that they could enable is beyond the scope of the present work, our results provide useful first indications.} Indeed, as we have shown, close asperities or thinner samples can significantly affect the individual microcontact behaviour, \REV{especially its stiffness}, and thus increase the diversity of available local responses, without requiring any significant change to the manufacturing method. However, such a widening of the design space comes with the challenge of using a more complex \REV{contact} model (e.g. full 3D finite elements like here) than the analytical one of Eqs.~\ref{Eq:A} and~\ref{Eq:P}. \REV{Similarly to design approaches devoted to volumetric structure problems, the interface inverse design process should therefore rely on combining such FEM models with optimization solvers (e.g.,\cite{djourachkovitch2021multiscale}) in order to better adapt the surface topography to the targeted performances.}

\section*{Acknowledgements}
The authors are indebted to the Carnot institute Ingénierie@Lyon, labelled by the French National Research Agency (ANR), for its support and funding. We thank Davy Dalmas for discussions.
\appendix
\section{Lists of asperity heights for the different metainterfaces (reference asperity placement)}
\label{sec:appendixA}

\begin{table*}[!h]
\centering
\caption{List of prescribed heights for the  metainterface with a friction law passing through three operating points (OP). Each operating point is reached by adding a new set of asperities with a common height.
"/" indicates that no asperity is present at this location.}
\resizebox{\columnwidth}{!}{%
\begin{tabular}{l|cccccccccccccccccccccc}
Asperity & 1 & 2 & 3 & 4 & 5 & 6 & 7 & 8 & 9 & 10 & 11 & 12 & 13 &  14 & 15 & 16\\
\hline
$h_i$ ($\mu$m) & 129.0 & 129.0 & 129.0 & / & 129.0 & 129.0 & 129.0 & 129.0 & 129.0 & 270.0 & 115.2 & / & 129.0 & 129.0 & 270.0 & 129.0\\
\hline
Asperity & 17 & 18 & 19 & 20 & 21 & 22 & 23 & 24 & 25 & 26 & 27 & 28 & 29 & 30 & 31 & 32\\
\hline
$h_i$ ($\mu$m) & 129.0 & 129.0 & 180.7 & 129.0 & 129.0 & 180.7 & 129.0 & 129.0 & 129.0 & 129.0 & 129.0 & 129.0 & 270.0 & 129.0 & / & 129.0\\
\hline
Asperity & 33 & 34 & 35 & 36 & 37 & 38 & 39 & 40 & 41 & 42 & 43 & 44 & 45 & 46 & 47 & 48\\
\hline
$h_i$ ($\mu$m) & / & 129.0 & 129.0 & 243.6 & 129.0 & 129.0 & 129.0 & 129.0 & 129.0 & 129.0 & 180.7 & 129.0 & 129.0 & 172.6 & 129.0 & / \\
\hline
Asperity & 49 & 50 & 51 & 52 & 53 & 54 & 55 & 56 & 57 & 58 & 59 & 60 & 61 & 62 & 63 & 64 \\
\hline
$h_i$ ($\mu$m) & 129.0 & 270.0 & 129.0 & 129.0 & 129.0 & 129.0 & 270.0 & 129.0 & 129.0 & 129.0 & / & 129.0 & 129.0 & 129.0 & 129.0 & 129.0 \\
\hline
\end{tabular}\label{Tab:HOP}
}
\end{table*}

\begin{table*}[!h]
\centering
\caption{List of prescribed heights for metainterfaces with a quasilinear (QL) friction law. QL1 and QL2 correspond, respectively, to the blue and red curves in Fig.~4 of~\cite{aymard2024designing}.
}
\resizebox{\columnwidth}{!}{%
\begin{tabular}{l|cccccccccccccccccccccc}
Asperity & 1 & 2 & 3 & 4 & 5 & 6 & 7 & 8 & 9 & 10 & 11 & 12 & 13 &  14 & 15 & 16\\
\hline
$h_i$ ($\mu$m) (QL1) & 101.0 & 83.0 & 95.0 & 113.0 & 89.0 & 83.0 & 95.0 & 83.0 & 89.0 & 101.0 & 89.0 & 89.0 & 101.0 & 89.0 & 113.0 & 83.0\\
$h_i$ ($\mu$m) (QL2) & 123.0 & 87.0 & 111.0 & 153.0 & 99.0 & 93.0 & 111.0 & 87.0 & 105.0 &  129.0 & 99.0 & 99.0 & 123.0 & 99.0 & 147.0 & 93.0\\
\hline
Asperity & 17 & 18 & 19 & 20 & 21 & 22 & 23 & 24 & 25 & 26 & 27 & 28 & 29 & 30 & 31 & 32\\
\hline
$h_i$ ($\mu$m) (QL1) & 101.0 & 107.0 & 83.0 & 137.0 & 83.0 & 95.0 & 131.0 & 95.0 & 95.0 & 125.0 & 83.0 & 113.0 & 83.0 & 83.0 & 89.0 & 95.0 \\
$h_i$ ($\mu$m) (QL2) & 123.0 & 141.0 & 87.0 & 189.0 & 87.0 & 117.0 & 177.0 & 117.0 & 117.0 & 171.0 & 87.0 & 141.0 & 87.0 & 93.0 & 105.0 & 111.0 \\
\hline
Asperity & 33 & 34 & 35 & 36 & 37 & 38 & 39 & 40 & 41 & 42 & 43 & 44 & 45 & 46 & 47 & 48\\
\hline
$h_i$ ($\mu$m) (QL1) & 89.0 & 101.0 & 107.0 & 125.0 &  83.0 & 101.0 & 89.0 & 119.0 & 89.0 & 83.0 & 89.0 & 119.0 & 83.0 & 83.0 & 95.0 & 113.0 \\
$h_i$ ($\mu$m) (QL2) & 105.0 & 117.0 & 135.0 & 165.0 & 87.0 & 129.0 & 99.0 & 159.0 & 105.0 & 93.0 & 105.0 & 159.0 & 93.0 & 93.0 & 111.0 & 147.0 \\
\hline
Asperity & 49 & 50 & 51 & 52 & 53 & 54 & 55 & 56 & 57 & 58 & 59 & 60 & 61 & 62 & 63 & 64 \\
\hline
$h_i$ ($\mu$m) (QL1) & 83.0 & 83.0 & 119.0 & 149.0 & 107.0 & 95.0 & 143.0 & 131.0 & 89.0 & 89.0 & 107.0 & 101.0 & 107.0 & 83.0 & 95.0 & 83.0 \\
$h_i$ ($\mu$m) (QL2) & 93.0 & 93.0 & 153.0 & 201.0 & 135.0 & 111.0 & 195.0 & 183.0 & 99.0 & 99.0 & 135.0 & 123.0 & 129.0 & 87.0 & 105.0 & 87.0 \\
\hline
\end{tabular}\label{Tab2}
}
\end{table*}


\begin{table*}[!h]
\centering
\caption{List of prescribed heights for metainterfaces with a bilinear (BL) friction law. BL1 and BL2 correspond, respectively, to the blue and red curves in Fig.~5 of~\cite{aymard2024designing}.}
\resizebox{\columnwidth}{!}{%
\begin{tabular}{l|cccccccccccccccccccccc}
Asperity & 1 & 2 & 3 & 4 & 5 & 6 & 7 & 8 & 9 & 10 & 11 & 12 & 13 &  14 & 15 & 16\\
\hline
$h_i$ ($\mu$m) (BL1) & 53.8 & 60.6 & 60.3 & 55.3 & 50.1 & 61.0 & 52.0 & 101.5 & 59.8 & 57.3 & 51.8 & 60.8 & 54.0 & 57.7 & 58.2 & 50.5\\
$h_i$ ($\mu$m) (BL2) & 63.3 & 62.4 & 58.7 & 60.9 & 56.0 & 57.6 & 55.5 & 54.3 & 60.6 & 55.8 & 56.3 & 55.0 & 94.5 & 64.9 & 58.9 & 59.8\\
\hline
Asperity & 17 & 18 & 19 & 20 & 21 & 22 & 23 & 24 & 25 & 26 & 27 & 28 & 29 & 30 & 31 & 32\\
\hline
$h_i$ ($\mu$m) (BL1) & 59.1 & 53.2 & 53.4 & 50.7 & 57.0 & 51.3 & 54.6 & 58.8 & 54.4 & 69.1 & 87.2 & 52.2 & 53.6 & 62.6 & 56.8 & 56.1 \\
$h_i$ ($\mu$m) (BL2) & 62.6 & 61.2 & 54.5 & 50.8 & 56.8 & 61.5 & 52.9 & 52.2 & 59.2 & 61.8 & 57.3 & 56.5 & 50.4 & 58.3 & 50.6 & 65.2 \\
\hline
Asperity & 33 & 34 & 35 & 36 & 37 & 38 & 39 & 40 & 41 & 42 & 43 & 44 & 45 & 46 & 47 & 48\\
\hline
$h_i$ ($\mu$m) (BL1) & 51.6 & 81.7 & 152.4 & 51.5 & 52.8 & 50.9 & 55.5 & 126.2 & 54.9 & 55.9 & 56.4 & 52.6 & 111.7 & 60.1 & 59.3 & 57.5 \\
$h_i$ ($\mu$m) (BL2) & 54.8 & 50.2 & 64.3 & 110.9 & 62.1 & 51.0 & 51.3 & 83.6 & 52.6 & 75.2 & 53.6 & 64.5 & 54.0 & 59.5 & 60.0 & 53.8 \\
\hline
Asperity & 49 & 50 & 51 & 52 & 53 & 54 & 55 & 56 & 57 & 58 & 59 & 60 & 61 & 62 & 63 & 64 \\
\hline
$h_i$ ($\mu$m) (BL1) & 93.6 & 50.3 & 59.6 & 57.9 & 55.7 & 56.6 & 52.4 & 51.1 & 77.0 & 55.1 & 54.2 & 53.0 & 58.4 & 58.6 & 65.7 & 72.8 \\
$h_i$ ($\mu$m) (BL2) & 145.4 & 51.9 & 60.3 & 63.9 & 51.7 & 63.0 & 52.4 & 51.5 & 55.3 & 53.3 & 57.8 & 57.0 & 63.6 & 53.1 & 68.4 & 58.1 \\
\hline
\end{tabular}\label{}
}
\end{table*}

\newpage
\section{Height lists of the variants used in the figures}
\label{sec:appendixB}

\begin{table*}[!h]
\centering
\caption{Table of heights for variants 1 and 2 of the OP metainterface.
"/" indicates that no asperity is present at this location.}
\resizebox{\columnwidth}{!}{%
\begin{tabular}{l|cccccccccccccccccccccc}
Asperity & 1 & 2 & 3 & 4 & 5 & 6 & 7 & 8 & 9 & 10 & 11 & 12 & 13 &  14 & 15 & 16\\
\hline
$h_i$ ($\mu$m) (variant 1) & 129.0 & 129.0 & 129.0 & / & 129.0 & 129.0 & 129.0 & 129.0 & 
                           129.0 & 129.0 & 115.2 & / & 129.0 & 129.0 & 243.6 & 129.0\\
$h_i$ ($\mu$m) (variant 2) & 129.0 & 129.0 & 129.0 & / & 129.0 & 129.0 & 129.0 & 129.0 & 
                           129.0 & 129.0 & 115.2 & / & 129.0 & 129.0 & 243.6 & 129.0\\
\hline
Asperity    & 17 & 18 & 19 & 20 & 21 & 22 & 23 & 24 & 25 & 26 & 27 & 28 & 29 & 30 & 31 & 32\\
\hline
$h_i$ ($\mu$m)  (variant 1)  & 129.0 & 129.0 & 180.7 & 129.0 & 129.0 & 180.7 & 129.0 & 129.0 
                           & 129.0 & 129.0 & 129.0 & 270.0 & 270.0 & 129.0 & / & 129.0 \\
$h_i$ ($\mu$m)  (variant 2)  & 129.0 & 129.0 & 180.7 & 129.0 & 129.0 & 180.7 & 129.0 & 129.0 
                           & 129.0 & 129.0 & 129.0 & 129.0 & 129.0 & 129.0 & / & 129.0 \\
\hline
Asperity & 33 & 34 & 35 & 36 & 37 & 38 & 39 & 40 & 41 & 42 & 43 & 44 & 45 & 46 & 47 & 48\\
\hline
$h_i$ ($\mu$m)   (variant 1)          & / & 129.0 & 129.0 & 270.0 & 270.0 & 129.0 & 129.0 & 129.0 
                          & 129.0 & 129.0 & 180.7 & 129.0 & 129.0 & 172.6 & 129.0 & / \\
$h_i$ ($\mu$m)   (variant 2)          & / & 129.0 & 129.0 & 129.0 & 129.0 & 129.0 & 129.0 & 129.0 
                          & 129.0 & 129.0 & 180.7 & 129.0 & 129.0 & 172.6 & 129.0 & / \\                          
                          
\hline
Asperity & 49 & 50 & 51 & 52 & 53 & 54 & 55 & 56 & 57 & 58 & 59 & 60 & 61 & 62 & 63 & 64 \\
\hline
$h_i$ ($\mu$m) (variant 1) & 129.0 & 129.0 & 129.0 & 129.0 & 129.0 & 129.0 & 270.0 & 129.0 
                         & 129.0 & 129.0 & / & 129.0 & 129.0 & 129.0 & 129.0 & 129.0 \\
$h_i$ ($\mu$m) (variant 2) & 129.0 & 129.0 & 129.0 & 270.0 & 270.0 & 129.0 & 270.0 & 129.0 
                         & 129.0 & 129.0 & / & 270.0 & 270.0 & 129.0 & 129.0 & 129.0 \\                         
\hline
\end{tabular}\label{Tab:HOP_var}
}
\end{table*}

\begin{table*}[!h]
\centering
\caption{Table of heights for variants 1 and 2 of the QL2 metainterface. 
}
\resizebox{\columnwidth}{!}{%
\begin{tabular}{l|cccccccccccccccccccccc}
Asperity & 1 & 2 & 3 & 4 & 5 & 6 & 7 & 8 & 9 & 10 & 11 & 12 & 13 &  14 & 15 & 16\\
\hline
$h_i$ ($\mu$m) (QL2 variant 1) & 93.0 & 93.0 & 93.0 & 177.0 & 87.0 & 87.0 & 87.0 & 135.0 & 195.0 &  189.0 & 105.0 & 105.0 & 171.0 & 99.0 & 99.0 & 99.0\\
$h_i$ ($\mu$m) (QL2 variant 2) & 93.0 & 93.0 & 93.0 & 93.0 & 87.0 & 87.0 & 87.0 & 87.0 & 105.0 &  105.0 & 105.0 & 105.0 & 99.0 & 99.0 & 99.0 & 99.0\\
\hline
Asperity & 17 & 18 & 19 & 20 & 21 & 22 & 23 & 24 & 25 & 26 & 27 & 28 & 29 & 30 & 31 & 32\\
\hline

$h_i$ ($\mu$m) (QL2 variant 1) & 105.0 & 111.0 & 111.0 & 111.0 & 111.0 & 165.0 & 147.0 & 105.0 
                              & 135.0 & 129.0 & 129.0 & 129.0 & 123.0 & 123.0 & 123.0 & 123.0 \\
$h_i$ ($\mu$m) (QL2 variant 2) & 117.0 & 117.0 & 117.0 & 117.0 & 111.0 & 111.0 & 111.0 & 111.0 
                              & 135.0 & 129.0 & 129.0 & 129.0 & 123.0 & 123.0 & 123.0 & 123.0 \\                              
\hline
Asperity & 33 & 34 & 35 & 36 & 37 & 38 & 39 & 40 & 41 & 42 & 43 & 44 & 45 & 46 & 47 & 48\\
\hline

$h_i$ ($\mu$m) (QL2 variant 1) & 153.0 & 99.0 & 141.0 & 117.0 & 135.0 & 147.0 & 87.0 & 93.0 
& 87.0 & 87.0 & 183.0 & 99.0 & 99.0 & 87.0 & 153.0 & 153.0 \\
$h_i$ ($\mu$m) (QL2 variant 2) & 153.0 & 153.0 & 147.0 & 147.0 & 141.0 & 141.0 & 135.0 & 135.0 
& 195.0 & 189.0 & 183.0 & 177.0 & 171.0 & 165.0 & 159.0 & 159.0 \\
\hline
Asperity & 49 & 50 & 51 & 52 & 53 & 54 & 55 & 56 & 57 & 58 & 59 & 60 & 61 & 62 & 63 & 64 \\
\hline

$h_i$ ($\mu$m) (QL2 variant 1) & 201.0 & 87.0 & 87.0 & 87.0 & 111.0 & 87.0 & 87.0 & 87.0 & 99.0 & 99.0 & 117.0 & 111.0 & 99.0 & 135.0 & 135.0 & 93.0 \\
$h_i$ ($\mu$m) (QL2 variant 2) & 93.0 & 93.0 & 87.0 & 87.0 & 87.0 & 87.0 & 87.0 & 201.0 & 111.0 & 105.0 & 105.0 & 99.0 & 99.0 & 99.0 & 93.0 & 93.0 \\
\hline
\end{tabular}\label{}
}
\end{table*}


\bibliographystyle{elsarticle-num} 
\bibliography{refs}
\end{document}